\documentclass[12pt]{article}

\usepackage{latexsym,amssymb,amsfonts,amsmath}
\usepackage[dvips]{graphicx}
\usepackage{setspace}

\renewcommand{\thefootnote}{\fnsymbol{footnote}}

\newcommand{\inst}{{\rm I}}
\newcommand{\ainst}{\bar{\rm I}}

\newcommand{\tr}{{\rm Tr}\,}
\newcommand{\trd}{{\rm Tr}_2\,}
\newcommand{\ttr}{\widetilde{\rm Tr}\,}
\newcommand{\T}{\mathbb{T}}
\newcommand{\W}{\mathbb{W}}

\newcommand{\I}{\mathbb{I}}

\newcommand{\la}{\langle}
\newcommand{\ra}{\rangle}
\newcommand{\qbar}{\bar{q}}

\newcommand{\para}{\parallel}

\newcommand{\z}{|\vec{z}_{\perp}|}
\newcommand{\fm}{{\rm fm}}

\newcounter{ridummy}

\newcommand{\newsection}[1]{
\vspace{5mm}
\pagebreak[3]
\setcounter{equation}{0}
\setcounter{subsection}{0}
\addtocounter{ridummy}{1}
\section{#1}
\nopagebreak
\medskip
\nopagebreak}

\newcommand{\mytitle}[1]{
\begin{center}
  \vspace*{1.0cm}
  \begin{doublespace}
    {\Large \bf #1}
  \end{doublespace}
  \vspace*{0.5cm}
  {\large Matteo Giordano\footnote{E--mail: matteo.giordano@df.unipi.it}
    and Enrico Meggiolaro\footnote{E--mail: enrico.meggiolaro@df.unipi.it} }\\
  \vspace*{0.5cm}{\normalsize
    {Dipartimento di Fisica, Universit\`a di Pisa,\\
      and INFN, Sezione di Pisa,\\
      Largo Pontecorvo 3,
      I--56127 Pisa, Italy.}}\\
\end{center}
}

\title{Instanton ef\mbox{}fects on Wilson--loop correlators: a new
      comparison  with numerical results from the lattice}
\author{Matteo Giordano and Enrico Meggiolaro}

\begin{document}
\thispagestyle{empty}
\setcounter{page}{0}

\noindent \hfill IFUP--TH/2009--17 \hspace{1cm}\\ \mbox{} \\
\mbox{} \hspace{1cm}  \hfill Revised version \hspace{1cm}\\ 
\mbox{} \hfill April 2010 \hspace{1cm}\\

\mytitle{Instanton ef\mbox{}fects on Wilson--loop correlators: a new
      comparison  with numerical results from the lattice}

\vspace*{\stretch{1}}

\abstract{Instanton ef\mbox{}fects on the Euclidean correlation function of
  two Wilson loops at an angle $\theta$, relevant to {\it soft} high--energy
  dipole--dipole scattering, are calculated in the Instanton Liquid Model and
  compared with the existing lattice data. Moreover, the instanton--induced
  dipole--dipole potential is obtained from the same correlation function at
  $\theta=0$, and compared with preliminary lattice data.} 

\vspace*{\stretch{1}}

\renewcommand{\thefootnote}{\arabic{footnote}}
\setcounter{footnote}{0}

\newsection{Introduction}

Since its discovery in 1975~\cite{BPST}, the instanton solution of the
Yang--Mills equations 
has been widely
studied, both in its mathematical properties and its phenomenological
applications~\cite{JR,CDG1,Poly,tHo,CDG2,ADHM,Shu}. Many insights
have been obtained in QCD through the use of instantons (see, e.g.,
the review~\cite{inst-rev} and references therein), and even if it is known
that they cannot provide the framework for a complete understanding of
strong interactions, it is interesting to investigate instanton
ef\mbox{}fects as they contribute to the nonperturbative dynamics.

Among the various open problems in QCD,  {\it soft} high--energy
hadron--hadron scattering is known to be of nonperturbative nature, and so it 
is worth studying what the consequences are of instantons on the 
scattering amplitudes. In the approach from the f\mbox{}irst principles of
QCD~\cite{Nachtmann91}, such amplitudes are related to properties of
the vacuum, namely the correlation functions of certain Wilson--line
and Wilson--loop operators. In particular, in the case of meson--meson
scattering, the scattering amplitudes can be reconstructed, after folding with
the appropriate mesonic wave functions, from the scattering amplitude of two
colour dipoles of f\mbox{}ixed transverse size; the latter are obtained 
from the correlation function of two rectangular Wilson loops, describing (in
the considered energy regime) the propagation of the colour
dipoles~\cite{DFK,Nachtmann97,BN,Dosch,LLCM1}. 
The physical, Minkowskian correlation functions can be
reconstructed from the ``corresponding'' Euclidean correlation
functions~\cite{Meggiolaro97,Meggiolaro98,Meggiolaro02,Meggiolaro05,crossing,Meggiolaro07,EMduality},
and so it has then been possible to investigate the problem of {\it soft}
high--energy scattering with some nonperturbative techniques available in
Euclidean Quantum Field Theory~\cite{instanton1,JP1,JP2,Janik,LLCM2,lattice}.

In this paper we shall derive a {\it quantitative} prediction of
instanton ef\mbox{}fects in the Euclidean loop--loop correlation function,
relevant to the problem of {\it soft} high--energy
scattering. This 
problem has already been addressed in Ref.~\cite{instanton1}, using
the so--called {\it Instanton Liquid Model} (ILM)~\cite{Shu}, but the
result reported in that paper contains a severe 
divergence, apparently not noticed by the authors. In this paper we
critically repeat the calculation, obtaining a well--def\mbox{}ined analytic
expression, which can be compared with the
lattice data presented and discussed in Ref.~\cite{lattice}. In this
paper we also calculate the instanton--induced { 
  dipole--dipole potential} from the correlation function of two
parallel Wilson loops~\cite{Pot}, and 
we compare the results with some preliminary data from the 
lattice.

The detailed plan of this paper is the following. 
In Section \ref{sec:inst_inst} we brief\mbox{}ly
recall the main features of instantons in Yang--Mills theory, 
we shortly describe the relevant aspects
of the ILM, and we discuss the method we will actually use in our
calculations, using as an example the expectation value of a
rectangular Wilson loop and the instanton--induced
$q\qbar$--potential~\cite{CDG2,DPP}. 
In Section \ref{sec:inst_theta}, after a brief review of how high--energy
scattering amplitudes can be reconstructed from the correlation function of
two rectangular Wilson loops at an angle $\theta$ in Euclidean space, we
evaluate this same correlation function and we critically compare the result
with the calculation of Ref.~\cite{instanton1}. We then compare our
quantitative prediction with the lattice results of Ref.~\cite{lattice}.
In Section \ref{sec:inst_dd} we calculate the correlation function of two 
parallel Wilson loops, from which we derive the instanton--induced
{dipole--dipole potential}, that we compare with some 
preliminary data from the lattice. 
Finally, in Section \ref{sec:inst_concl} we draw our conclusions. 
Some technical details are discussed in the Appendices\footnote{In this 
  paper we will deal only with the Euclidean theory, and so we will
  understand that the metric is Euclidean and that expectation values
  have to be taken with respect to the Euclidean functional
  integral, except where explicitly stated. }.

\newsection{Wilson loops in the ILM}
\label{sec:inst_inst}

Instantons (anti--instantons) are self--dual
(anti--self--dual) solutions with f\mbox{}inite action of the
classical Yang--Mills equations of motion in Euclidean
space~\cite{BPST}. 
We will be interested only in the solutions with {topological
  charge} $q=+1$ and
$q=-1$, which we will refer to as the instanton ($\inst$) and the
anti--instanton ($\ainst$), respectively. In the case of $SU(2)$
Yang--Mills theory, the instanton solution in regular covariant (Lorenz)
gauge reads, for the standard colour orientation,
\begin{equation}
  \label{eq:instanton-sol}
  A_\mu^{a}(x;z,\inst) = -\frac{2}{g}\frac{\eta_{a\mu\nu}(x-z)_\nu}{(x-z)^2 +
  \rho^2},\qquad A_\mu =A_\mu^{a}\frac{\sigma^a}{2},\,\,a=1,2,3,
\end{equation}
where $z_\mu$ and $\rho$ are free parameters that correspond to the
position of the center and the radius of the instanton, respectively,
$\eta_{a\mu\nu}$ are the self--dual \mbox{'t Hooft}
symbols~\cite{tHo}, 
and $\sigma^a$ are the Pauli matrices. The anti--instanton
solution $A_\mu^{a}(x;z,\ainst)$ is obtained by replacing
$\eta_{a\mu\nu}\to\bar{\eta}_{a\mu\nu}$ in
Eq.~(\ref{eq:instanton-sol}), where the latter are the 
anti--self--dual 't Hooft symbols~\cite{tHo}. 
Applying a $SU(2)$ transformation on Eq.~(\ref{eq:instanton-sol}),
$A_\mu \to UA_\mu U^\dag$ with $U\in SU(2)$, we obtain 
another solution with a dif\mbox{}ferent colour orientation.
Instantons in $SU(N_c)$ Yang--Mills theory can be obtained by embedding the
$SU(2)$ solution (\ref{eq:instanton-sol}) in a $SU(2)$ subgroup
of $SU(N_c)$, and it has been shown that these are the only solutions
with $|q|=1$~\cite{ADHM}. We will take the standard--oriented solution
to be the embedding of the (standard) $SU(2)$ solution in the
upper--left $2\times 2$ corner of an $N_c\times N_c$ matrix; 
the other solutions are obtained by applying an $SU(N_c)$ colour
rotation on this one. 

Early attempts at a description of the QCD vacuum in terms of an
ensemble of instantons and anti--instantons adopted the picture of a
{\it dilute gas}~\cite{CDG2}, but suf\mbox{}fered from severe infrared divergences
due to the presence of large--size instantons. The proposal
of~\cite{Shu} was instead that the pseudoparticles form a {\it dilute
liquid}: the diluteness of the medium allows for a meaningful
description in terms of individual pseudoparticles, while the infrared
problem is solved assuming that interactions stabilise the instanton
radius. 
In particular, this model assumes that
the vacuum can be described as a liquid of instantons and
anti--instantons, with equal densities $n_{\inst} =
n_{\ainst} = n/2$, and with radius distributions strongly peaked
around the same average value: in practice, one performs the
calculations using a $\delta$--like distribution, i.e., f\mbox{}ixing the
radii of the pseudoparticles to a given value $\rho$. Using the SVZ
sum rules~\cite{SVZ,Narison} to 
relate vacuum properties with experimental quantities, the
phenomenological values for the total density and the average radius
are estimated to be~\cite{inst-rev} $n\simeq 1 {\rm \,fm}^{-4}$, $\rho \simeq
1/3 {\rm\, fm}$, which lead to a small {\it packing fraction} $nV_1
\sim 1/20$ [here $V_1= (\pi^2/2) \rho^4$ is the volume of a 3--sphere
of radius $\rho$, which is approximately the space--time volume occupied by a
single pseudoparticle]; this indicates that the instanton liquid is indeed
fairly dilute\footnote{These values and the corresponding picture are obtained
in the realistic case of three light--quark f\mbox{}lavours ($u$, $d$, $s$):
{when comparing the analytic results with the numerical
lattice calculations, 
we have to take into account that the 
latter are performed} in the {\it quenched} approximation of QCD.}. We refer
the interested reader to~\cite{inst-rev} and references therein for a
more detailed discussion of the model, both on its theoretical
foundations~\cite{IM-P,DP} and on the development of numerical methods
of investigation~\cite{Shu2}. In the rest of this paper we will make
use only of the general picture described above to give estimates of
the {relevant quantities}. 

To calculate the expectation value of an operator $\cal O$ one needs
to average over the ensemble with some density function ${\cal
  D}={\cal D}(\{z\},\{U\})$, where $\{z\}$ denotes collectively the
coordinates of the pseudoparticles and $\{U\}$ their orientations 
in colour space; in principle, such a function should  be determined
by properly taking instanton interactions into account. 
Starting with an ensemble of $N_{\inst}$ instantons and $N_{\ainst}$
anti--instantons\footnote{For generality, we keep $N_{\inst}$ and $N_{\ainst}$
  f\mbox{}ixed but independent in the calculation; we will
  set $N_{\inst}=N_{\ainst}$ when needed.} in a space--time
volume $V$, we obtain the desired result by taking the
``thermodynamic limit'' $N_{\inst},N_{\ainst},V\to\infty$, while keeping the
densities $n_{\inst} \equiv N_{\inst}/V$ and $n_{\ainst} \equiv
N_{\ainst}/V$ f\mbox{}ixed: 
\begin{align}
  \langle {\cal O} \rangle_{\rm ILM} \equiv \lim_{\substack{N_{\inst,\ainst},V\to\infty,\\
  n_{\inst,\ainst}\,{\rm f\mbox{}ixed}}}{\displaystyle \int d^{4N}z\,d^N U\, 
  {\cal D}(\{z\},\{U\})\, {\cal O}(\{z\},\{U\})}.
\end{align}
Here ${\cal O}(\{z\},\{U\})$ denotes the operator $\cal O$ evaluated in
the f\mbox{}ield conf\mbox{}iguration generated by the pseudoparticles, and $dU$ is
the normalised Haar measure, $\int dU =1$.
Since the medium is dilute, the f\mbox{}ield conf\mbox{}iguration is
approximately equal to the superposition of single (anti--)instanton
f\mbox{}ields, 
\begin{equation}
\label{eq:liquidconf}
  A_{\mu}(x) = \sum_{i=1}^{N} U(i)A_\mu(x;z_i,\sigma_i)U^\dag(i),
\end{equation}
where $N=N_{\inst}+N_{\ainst}$ is the total pseudoparticle number,
$z_i$ and $U(i)$ are the position and colour orientation of
pseudoparticle $i$, respectively, and $\sigma_i=\inst,\ainst$
indicates if pseudoparticle $i$ is an instanton or an anti--instanton.

To perform calculations, we take an approximate density function that
describes the instanton liquid as an ensemble of free
particles with some (unspecif\mbox{}ied) strong core repulsion, which
approximately localises the instantons in distinct ``cells'' of volume
$V_{\rm cell}$, i.e., 
\begin{equation}
  \label{eq:density}
  {\cal D}(\{z\}) = {\cal N}^{-1}\sum_{P} \prod_{j=1}^{N}
  \chi_{D_j}(z_{P(j)}),\qquad {\cal N}= N!\,V_{\rm cell}^N
\end{equation}
where $\{D_j\}$ is a partition in cells of the space--time region $D$
occupied by the ensemble, $\cup_j D_j = D$, $\chi_{D_j}$ 
is the characteristic function of $D_j$, $P$ denotes a permutation of
$\{1,\ldots,N\}$, and ${\cal N}$ is the normalisation; the volume of
each cell is $\int d^4z\, \chi_{D_j}(z) = V_{\rm 
  cell}$. The coordinates of pseudoparticle $j$ are denoted as $z_j$,
and we take the f\mbox{}irst $N_{\inst}$ pseudoparticles to be instantons, and the
remaining $N_{\ainst}$ to be anti--instantons,
\begin{equation}
  \sigma_k=\left\{
  \begin{aligned}
      &\inst,  &&1\le k\le N_{\inst},\\
      &\ainst,  &&N_{\inst}+1 \le k \le N.
  \end{aligned}
\right.
\end{equation}
The function (\ref{eq:density}) is by no means a precise description of the instanton
ensemble: it has the purpose of capturing the main features of the liquid
picture, namely the diluteness and the uniform density of the medium, while at
the same time keeping the calculation feasible. Notice that ${\cal D}$ does not
depend on the colour orientation of the pseudoparticles. 
The precise shape of the cells should not be important, and one can
choose it according to the geometry of the problem. 

For a local operator ${\cal O}_{\rm local}(x)$, only
conf\mbox{}igurations with a single pseudoparticle located near the
point $x$ have both non--negligible ef\mbox{}fects and non--negligible
measure in conf\mbox{}iguration space, due to the diluteness
of the medium and to the short--range nature of instanton
ef\mbox{}fects. The evaluation of expectation values is more
complicated for non--local operators such as Wilson lines and Wilson
loops, since they involve the value of the f\mbox{}ields on a curve ${\cal C}$.
However, by properly subdividing ${\cal C}$ into small segments, one can
exploit again the diluteness of the medium and the short--range
nature of instanton ef\mbox{}fects, in order to consider each segment to
be af\mbox{}fected only by the f\mbox{}ield of a single pseudoparticle.

\begin{figure}[t]
  \centering
  \rotatebox{90}{\includegraphics[width=0.3\textwidth]{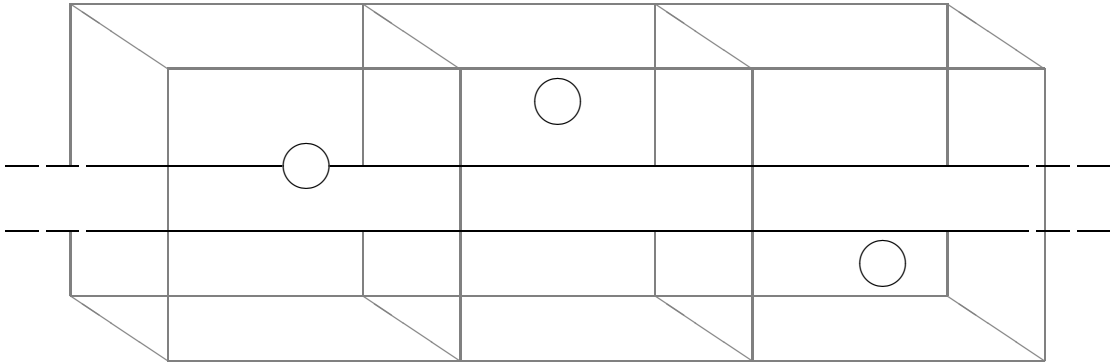}}
  \caption[Schematic representation of a Wilson loop in the instanton
    liquid.]{Schematic representation of a Wilson loop in the instanton
    liquid. Circles represent pseudoparticles with the same radius
    $\rho$, each lying in a cell.}
  \label{fig:instloop}
\end{figure}

As an example, consider the case of the expectation value of a
rectangular Wilson loop ${\cal W}$ {of temporal extension $T$ and
spatial width $R$ (say, in the $1$ direction)}.
As it is well known, one can extract from this expectation value the
static quark--antiquark  potential ${\cal V}_{q\qbar}$ 
using the relation~\cite{Wil,BW}
\begin{equation}
  \la {\cal W} \ra = \frac{1}{N_c}\tr
   [{\rm W}^{(-)\dag}{\rm W}^{(+)}] \mathop{=}_{T\to\infty} {\cal K}e^{-T{\cal V}_{q\qbar}(R)},
\end{equation}
where we have neglected the effect of the transverse ``links''
{connecting the Wilson lines ${\rm W}^{(\pm)}$, which describe the
propagation of the quark and of the antiquark}, and where ${\cal K}$
is some proportionality constant.  
The effect of a single instanton located at $z$ on an infinite Wilson
loop, which we denote with $w(z,\inst)$, 
is given by the following expression~\cite{CDG2},
\begin{equation}
   \label{eq:loop-one-inst}
   w(z,\inst) \equiv
   1-\frac{2}{N_c}\Delta(z),
\end{equation}
where 
\begin{equation}
\label{eq:alpha}
 \begin{aligned}
   &\Delta(z) = 1-\cos\alpha_+\cos\alpha_- -
   \hat{n}_+\cdot\hat{n}_-\sin\alpha_+\sin\alpha_-\,,\\
   &\alpha_\pm = \frac{\pi\|n_\pm\|}{\sqrt{\|n_\pm\|^2 + \rho^2}}\,,
   \qquad n^a_{\pm}=-(\pm \frac{R}{2}\delta_{1a}-z_a)\,.
 \end{aligned}
 \end{equation}
The result does not depend on the colour orientation of the instanton,
and it does not change if we replace the instanton with an
anti--instanton. In order to calculate the effect of the instanton
liquid, i.e., of the whole ensemble, on the Wilson loop, it is natural to see it 
as the result of a time series of
interactions between the colour dipole and the pseudoparticles: this
suggests dividing space--time into time--slices of thickness $\delta T$, and
then each slice into cells of volume $V_{\rm cell} = V_{\rm cell}^{(3)}\,\delta
T$. {The situation is illustrated in Fig.~\ref{fig:instloop}}. Here
$\delta T$ and $V_{\rm cell}^{(3)}$ are determined  
by the following requirements:
\begin{enumerate}
\item only one pseudoparticle falls in $V_{\rm cell}^{(3)}\,\delta T $ on
  the average, i.e., $nV_{\rm cell}=1$;
\item the spatial size $V_{\rm cell}^{(3)}$ is large enough to accommodate
  the whole dipole, and also large enough so that the pseudoparticles which 
  fall outside of the cell where the dipole lies will not af\mbox{}fect it, 
  being too distant;
\item $\delta T$ is large enough for an (anti--)instanton to fully
  af\mbox{}fect the portions of the two Wilson lines that fall in $V_{\rm cell}$,
  i.e., $\delta T$ can be considered inf\mbox{}inite ($\delta T\gg \rho$) as
  far as the colour rotation induced on the Wilson lines by the
  pseudoparticle is concerned, and so one can use the colour rotation
  angles $\alpha_\pm$ given in Eq.~(\ref{eq:alpha}).
\end{enumerate}
The region of 3--space where an (anti--)instanton can af\mbox{}fect the operator is
approximately determined by requiring that the pseudoparticle is not too
far from the two long sides (say, more than $\rho$), which
implies 
\begin{equation}
   V_{\rm cell}^{(3)} \sim (R + 2\rho)(2\rho)^2;
\end{equation}
in this way the two segments of the loop lying in a given slice fall
in the same cell. With this choice we fulf\mbox{}ill condition (2.); imposing
condition (1.) we obtain
\begin{equation}
  \frac{\delta T}{\rho} \sim \frac{81\rho}{4(R + 2\rho)} \sim
  \frac{10}{1+\frac{R}{2\rho}}, 
\end{equation}
which is reliable for $R\lesssim 2\rho$, so that $\delta
T/\rho \gtrsim 5$ in this regime, and condition (3.) is satisf\mbox{}ied. 
For larger dipoles, the quark and the antiquark are more likely to interact
with distinct instantons: the expectation value should then factorise
for large $R$.  

At this point one can perform the average over the instanton ensemble:
since the procedure is the same as the one adopted in~\cite{CDG2}, we will not
report the calculation here, and we simply quote the result which
will be used in the following: 
\begin{equation}
\label{eq:ILM-qqbarpot}
  \la{\cal W}\ra_{\rm ILM} = 
e^{-T{\cal V}_{q\qbar}(R)}, \quad {\cal V}_{q\qbar}(R) = n\frac{2}{N_c}\int d^{3}z \, \Delta(z).
\end{equation}
This result is the same as the one found in Ref.~\cite{DPP} with
dif\mbox{}ferent, more sophisticated methods. 
As $\Delta(z)$ is localised around the position in
3--space of the quark and the antiquark, for large $R$ the
Wilson--loop expectation value factorises, and the potential becomes the sum
of two constant and equal terms, which are interpreted as the renormalisation
of the quark mass~\cite{DPP}.

\newsection{High--energy meson--meson scattering and the Wilson--loop
  correlation functions in the ILM} 
\label{sec:inst_theta}

As has already been recalled in the Introduction, in the {\it soft}
high--energy regime the mesonic scattering amplitudes can be reconstructed
from the correlation function of two Euclidean Wilson loops. In this Section
we give f\mbox{}irst a brief account, for the benef\mbox{}it of the reader, of the 
main points of the functional--integral  approach to the problem of elastic 
meson--meson scattering, referring the interested reader to the original
papers~\cite{DFK,Nachtmann97,BN,Dosch,LLCM1} and to the
book~\cite{pomeron-book}. We shall use the same notation adopted in
Ref.~\cite{lattice}, where a more detailed presentation can be
found\footnote{Since no ambiguity can arise, with respect to
  Ref.~\cite{lattice} we drop the subscript $E$ and ``tildes'' from 
  Euclidean  quantities in order to avoid a cumbersome notation.}. We then 
critically repeat the calculation of the relevant correlation function in the
ILM, and compare the result with the one found in Ref.~\cite{instanton1}.  

The elastic scattering amplitudes of two mesons (taken for simplicity 
with the same mass $m$) in the {\it soft} high--energy regime can be
reconstructed in two steps. One f\mbox{}irst evaluates the scattering amplitude 
of two $q\qbar$ colour dipoles of f\mbox{}ixed transverse sizes $\vec{R}_{1\perp}$
and $\vec{R}_{2\perp}$, and f\mbox{}ixed longitudinal momentum fractions $f_1$ and
$f_2$ of the two quarks in the two dipoles, respectively; the mesonic
amplitudes are then obtained after folding the dipoles' amplitudes
with the appropriate squared wave functions, describing the
interacting mesons. The dipole--dipole amplitudes are given by the
2--dimensional Fourier transform, with respect to the transverse
distance $\vec{z}_{\perp}$, of the normalised (connected) correlation
function of two rectangular Wilson loops,
\begin{equation}
  {\cal M}_{(dd)} (s,t;\vec{R}_{1\perp},f_1,\vec{R}_{2\perp},f_2) \equiv
-i~2s \displaystyle\int d^2 \vec{z}_\perp
e^{i \vec{q}_\perp \cdot \vec{z}_\perp}
{\cal C}_M(\chi;\vec{z}_\perp;1,2) ,
\label{scatt-loop}
\end{equation}
where the arguments ``$1$'' and ``$2$'' stand for ``$\vec{R}_{1\perp},
f_1$'' and ``$\vec{R}_{2\perp}, f_2$'' respectively, $t =
-|\vec{q}_\perp|^2$ ($\vec{q}_\perp$ being the transferred momentum)
and $s=2m^2(1+\cosh\chi)$.
The correlation function ${\cal C}_M$ is def\mbox{}ined as the
limit $\displaystyle {\cal C}_M \equiv \lim_{T\to\infty} {\cal G}_M $
of the correlation function of two loops of f\mbox{}inite length $2T$,
\begin{equation}
  {\cal G}_M(\chi;T;\vec{z}_\perp;1,2) \equiv
\dfrac{ \langle {\cal W}^{(T)}_1 {\cal W}^{(T)}_2 \rangle_M }{
\langle {\cal W}^{(T)}_1 \rangle_M
\langle {\cal W}^{(T)}_2 \rangle_M } - 1,
\label{GM}
\end{equation}
where $\langle\ldots\rangle_M$ are averages in the sense of the QCD
(Minkowskian) functional integral. Here ${\cal W}^{(T)}_{1,2}$ are Minkowskian
Wilson loops evaluated along the paths ${\cal C}_{1,2}$, made up of the
classical trajectories of the quarks and antiquarks inside the two mesons, and
closed by straight--line paths in the transverse plane at proper times $\pm
T$.  
The partons' trajectories form a hyperbolic angle $\chi$ in the longitudinal
plane, and they are located at  $(1-f_i) \vec{R}_{i\perp}$ (quark) and $-f_i
\vec{R}_{i\perp}$ (antiquark), $i=1,2$, in the transverse plane.

The Euclidean counterpart of Eq.~(\ref{GM}) is
\begin{equation}
  {\cal G}_E(\theta;T;\vec{z}_\perp;1,2) \equiv
\dfrac{ \langle {\cal W}^{(T)}_1 {\cal W}^{(T)}_2 \rangle_E}
{\langle {\cal W}^{(T)}_1 \rangle_E
\langle {\cal W}^{(T)}_2 \rangle_E } - 1,
\label{GE}
\end{equation}
where now $\langle\ldots\rangle_E$ is the average in the sense of the
Euclidean QCD functional integral. With a little abuse of notation, we denote
with the same symbol ${\cal W}^{(T)}_{1,2}$ the Euclidean Wilson loops 
calculated on the following straight--line paths\footnote{The
  fourth Euclidean coordinate $X_{4}$ is taken to be the ``Euclidean
  time''.},
\begin{align}
  {\cal C}_1 :& \quad
X^{1q}_{\mu}(\tau) = z_{\mu} + u_{1\mu}  \tau
+ (1-f_1) R_{1\mu}, ~~~~ 
X^{1\bar{q}}_{\mu}(\tau) =
z_{\mu} + u_{1\mu} \tau  
- f_1 R_{1\mu} ,
\nonumber \\
{\cal C}_2 :& \quad
X^{2q}_{\mu}(\tau) = u_{2\mu}  \tau + (1-f_2) R_{2\mu},
~~~~
X^{2\bar{q}}_{\mu}(\tau) =
        u_{2\mu}  \tau - f_2 R_{2\mu} , 
\label{trajE}
\end{align}
with $\tau\in [-T,T]$, and closed by straight--line paths in the
transverse plane at $\tau=\pm T$. Here
\begin{eqnarray}
&u_{1} =
 \left(  \sin\frac{\theta}{2}, \vec{0}_\perp, \cos\frac{\theta}{2} \right) ,~~~
u_{2} =
 \left( -\sin\frac{\theta}{2}, \vec{0}_\perp, \cos\frac{\theta}{2} \right),
\label{p1p2E}
\end{eqnarray}
and $R_{1} = (0,\vec{R}_{1\perp},0)$, $R_{2} =
(0,\vec{R}_{2\perp},0)$, $z = (0,\vec{z}_\perp,0)$. 
Again, we def\mbox{}ine the correlation function with the IR cutof\mbox{}f $T$ removed 
as $\displaystyle {\cal C}_E \equiv \lim_{T\to\infty} {\cal G}_E $. 

\begin{figure}[t]
  \centering
  {\includegraphics[width=0.6\textwidth]{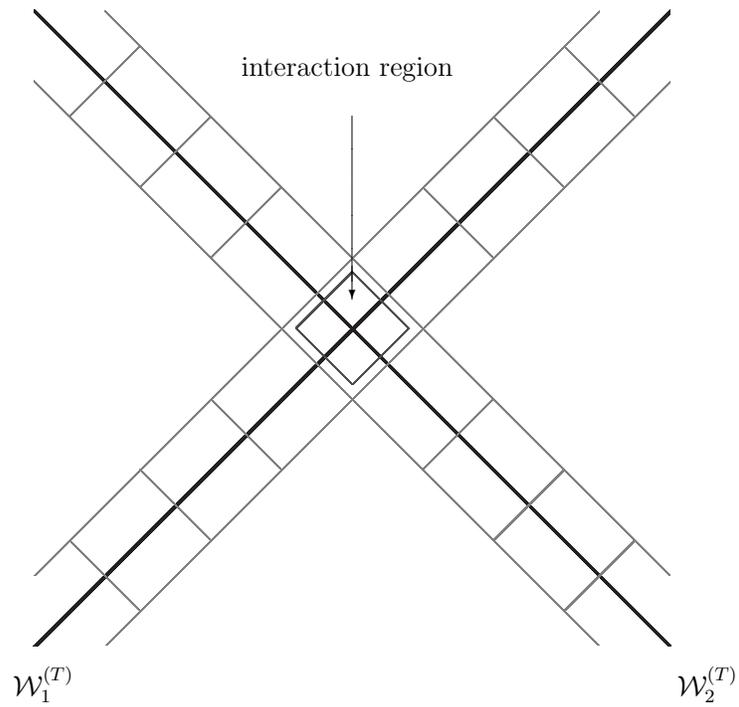}}
  \caption{Longitudinal projection of the two loops, showing the
    partition in cells relevant to the case of loops at an angle $\theta$.}
  \label{fig:intreg}
\end{figure}

It has been shown
in~\cite{Meggiolaro02,Meggiolaro05,EMduality} that  
the correlation functions in the two theories are connected by the
{\it analytic--continuation relations\/}\footnote{The functions on the
  right--hand side of Eqs.~(\ref{analytic}) and (\ref{analytic_C})  are  
  understood as the {\it analytic extensions} of the
  Euclidean and Min\-kow\-skian correlation functions, starting from the
  real intervals  $(0,\pi)$ and $\mathbb{R}^+$ of the respective angular
  variables, with positive real $T$ in both cases, into domains of the complex
  variables $\theta$ (resp.~$\chi$) and $T$ in a two--dimensional complex
  space. See Ref.~\cite{EMduality} for a more detailed discussion.}
\begin{eqnarray}
{\cal G}_M(\chi;T;\vec{z}_\perp;1,2)
&=& {\cal G}_E (-i\chi;iT;\vec{z}_\perp;1,2) ,
\qquad \forall\,\chi\in \mathbb{R}^+, \nonumber\\
  {\cal G}_E(\theta;T;\vec{z}_\perp;1,2)
&=& {\cal G}_M (i\theta;-iT;\vec{z}_\perp;1,2) ,
\qquad \forall\,\theta\in (0,\pi).
\label{analytic}
\end{eqnarray}
Under certain analyticity hypotheses in the $T$ variable, 
the following relations are obtained for the
correlation functions with the IR cutof\mbox{}f $T$
removed~\cite{Meggiolaro05,EMduality}: 
\begin{eqnarray}
{\cal C}_M(\chi;\vec{z}_\perp;1,2) &=&
{\cal C}_E(-i\chi;\vec{z}_\perp;1,2) ,
\qquad \forall\,\chi\in \mathbb{R}^+,\nonumber \\
  {\cal C}_E(\theta;\vec{z}_\perp;1,2) &=&
{\cal C}_M(i\theta;\vec{z}_\perp;1,2) ,
\qquad \phantom{-}\forall\,\theta\in (0,\pi).
\label{analytic_C}
\end{eqnarray}
We turn now to the calculation of instanton ef\mbox{}fects on the correlation
function ${\cal C}_E$. As it is explained in Appendix \ref{app:longmom}, we
can set $f_1=f_2=1/2$ without loss of generality; we also drop the dependence
on $f_i$ in the following formulas, i.e., we understand ${\cal
  C}_E(\theta;\vec{z}_\perp;\vec{R}_{1\perp},\vec{R}_{2\perp}) \equiv {\cal
  C}_E(\theta;\vec{z}_\perp;\vec{R}_{1\perp},f_1=1/2,\vec{R}_{2\perp},f_2=1/2)$. 
Neglecting 
the transverse links connecting the quark and antiquark trajectories, the Wilson 
loops are written as 
\begin{equation}
  {\cal W}_i^{(T)} = \frac{1}{N_c}\tr [
  {\rm W}^{(\qbar)\dag}_i{\rm W}^{(q)}_i] , \qquad i=1,2,
\end{equation}
with
\begin{align}
  {\rm W}^{(\qbar)}_i &= {\cal P} \exp\left[ -ig \displaystyle\int_{-T}^{T}
      A_{\mu}(X^{i\qbar}(\tau)){u_{i\mu}} d\tau \right], \qquad i=1,2
\end{align}
and similarly for ${\rm W}^{(q)}_i$.
Since the (long) sides of the two loops have dif\mbox{}ferent directions in the
longitudinal plane, their relative distance grows as we move away from
the center of the conf\mbox{}iguration; as a consequence, only
pseudoparticles falling in a f\mbox{}inite interaction region have ef\mbox{}fects on
both loops (see Fig.~\ref{fig:intreg}), and will therefore contribute
to the connected part of the loop--loop correlator. This region is roughly
determined by requiring that the distance of the pseudoparticles from
both loops in the transverse and in the longitudinal plane does not
exceed $\sim \rho$. Then $V_{\rm int} = V_\parallel V_\perp$, where
$\para$ and $\perp$ refer to the longitudinal and transverse planes,
respectively, with  
\begin{align}
  V_\perp &\sim \left(|\vec{z}_{\perp}| + \frac{|\vec{R}_{1\perp}|+|\vec{R}_{2\perp}|}{2}
   + 2\rho\right) \left(\frac{|\vec{R}_{1\perp}|+|\vec{R}_{2\perp}|}{2} + 2\rho\right),\\
   V_\parallel &\sim \frac{(2\rho)^2}{\sin\theta}.
\end{align}
We now make the following approximation, considering only one
instanton or anti--instanton in the interaction region, all the other
pseudoparticles interacting with one loop only (or not interacting at
all with the loops). Clearly, $V_\parallel$ blows up as $\theta\to
0,\pi$, and so does the number of pseudoparticles in the interaction 
region $nV_{\rm int} \propto n\rho^4/\sin\theta$; thus, the {\it one--instanton
approximation} can be good if the angle is not too close to $0,\pi$. In
this case we can perform the integration over the colour degrees of
freedom independently for the two loops, except for the pseudoparticle
in the interaction region, but since this last integration is trivial
we obtain simply
\begin{align}
  {\cal W}_1^{(T)} {\cal W}_2^{(T)} \to \left[\prod{}^{'} w_1(k)\prod{}^{'} w_2(k)\right]
  w_1(0)w_2(0),
\end{align}
where the prime indicates that we exclude the terms corresponding to
the interaction region, and $0$ refers to the interacting
pseudoparticle. 
Here $w_i(k) = 1 - (2/N_c) \Delta_i(x_{P(k)})$, with $x_{P(k)}$ being
the position of the pseudoparticle lying in the $k$--th cell [see
Eq.~(\ref{eq:loop-one-inst})], and with  
\begin{align}
    \Delta_i(x) &= 1-\cos\alpha_{i+}\cos\alpha_{i-} -
  \hat{n}_{i+}\cdot\hat{n}_{i-}\sin\alpha_{i+}\sin\alpha_{i-},\nonumber\\
  \alpha_{i\pm} &= \frac{\pi\|n_{i\pm}\|}{\sqrt{\|n_{i\pm}\|^2 +
      \rho^2}},\nonumber\\
n^a_{1\pm} &= \eta_{a\mu\nu}u_{1\mu}(z \pm \frac{R_1}{2}-x)_\nu, \quad
n^a_{2\pm} = \eta_{a\mu\nu}u_{2\mu}(\pm \frac{R_2}{2}-x)_\nu.
\end{align}
Performing the integration over the pseudoparticle positions,
making use of formula (\ref{eq:inst-av-over-pos-2}) in Appendix
\ref{app:C}, and dividing by the 
expectation values of the loops [see Eq.~(\ref{eq:ILM-qqbarpot})] we
f\mbox{}inally obtain for the {\it normalised connected correlation
  function} with $T\to\infty$ 
\begin{multline}
  {\cal
    C}_E^{\rm
    (ILM)}(\theta;\vec{z}_\perp;\vec{R}_{1\perp},\vec{R}_{2\perp})\\=  
  \dfrac{\displaystyle 1 + n\int_{\rm int} d^4x
    \left[\left(1-\frac{2}{N_c}\Delta_1(x)\right)
      \left(1-\frac{2}{N_c}\Delta_2(x)\right) -1\right]}{\displaystyle 
    \left[1-n\int_{\rm int} d^4x\,
      \frac{2}{N_c}\Delta_1(x)\right]\left[1-n\int_{\rm int} d^4x\,
      \frac{2}{N_c}\Delta_2(x)\right]}   -1, 
\end{multline}
where ``int'' indicates that the integration range is restricted to
the interaction region, since all the other terms cancel between the
numerator and denominator. Expanding to f\mbox{}irst order in $n$, for
consistency with the one--instanton approximation, we f\mbox{}inally get
\begin{equation}
  {\cal
    C}_E^{\rm (ILM)}(\theta;\vec{z}_\perp;\vec{R}_{1\perp},\vec{R}_{2\perp}) =
  n\left(\frac{2}{N_c}\right)^2\int d^4x\,
  \Delta_1(x)\Delta_2(x), 
\end{equation}
having extended the integration range to the whole space--time since the
integrand is now rapidly vanishing. Exploiting the properties of the
't Hooft symbols we f\mbox{}ind 
\begin{align}
\label{eq:ILMthetaquant}
&\begin{aligned}
  \|n_{1\pm}\|^2 &= \left(\vec{z}_{\perp}\pm \frac{\vec{R}_{1\perp}}{2} - \vec{x}_{\perp}\right)^2 + \left(\sin\left(\frac{\theta}{2}\right)
  x_4 - \cos\left(\frac{\theta}{2}\right) x_1\right)^2, \\
  \|n_{2\pm}\|^2 &= \left(\pm \frac{\vec{R}_{2\perp}}{2} - \vec{x}_{\perp}\right)^2 + \left(\sin\left(\frac{\theta}{2}\right)
  x_4 + \cos\left(\frac{\theta}{2}\right) x_1\right)^2, \\
\end{aligned}
\intertext{and, moreover,}
\label{eq:ILMthetaquant2}
&\begin{aligned}
  n_{1+}\cdot n_{1-} &= 
\left(\vec{z}_{\perp}  - \vec{x}_{\perp}\right)^2  - \left(\frac{\vec{R}_{1\perp}}{2}\right)^2
 + \left(\sin\left(\frac{\theta}{2}\right) x_4 - \cos\left(\frac{\theta}{2}\right) 
  x_1\right)^2, \\ 
  n_{2+}\cdot n_{2-} &= \vec{x}_{\perp}^{\,2} - \left(\frac{\vec{R}_{2\perp}}{2}\right)^2
  + \left(\sin\left(\frac{\theta}{2}\right) x_4 + \cos\left(\frac{\theta}{2}\right) x_1\right)^2.
\end{aligned}
\end{align}
We can then make the dependence on $\theta$ explicit by performing a
change of variables:
\begin{align}
\left\{
\begin{aligned}
  x_4' &= \sin\left(\frac{\theta}{2}\right) x_4 - \cos\left(\frac{\theta}{2}\right) x_1,\\
  x_1' &= \sin\left(\frac{\theta}{2}\right) x_4 + \cos\left(\frac{\theta}{2}\right) x_1,
\end{aligned}\right.
\end{align}
which f\mbox{}inally leads to\footnote{We drop the absolute value from the
  Jacobian $|\sin\theta|$, since we are limiting to
  $\theta\in (0,\pi)$.}
\begin{equation}
  \label{eq:C_inst}
{\cal C}_E^{\rm (ILM)}(\theta;\vec{z}_\perp;\vec{R}_{1\perp},\vec{R}_{2\perp}) \\
= n\left(\frac{2}{N_c}\right)^2\frac{1}{\sin\theta}\,F(\vec{z}_{\perp},\vec{R}_{1\perp},\vec{R}_{2\perp}),
  \end{equation}
with
\begin{equation}
  \label{eq:inst_int_num}
  F(\vec{z}_{\perp},\vec{R}_{1\perp},\vec{R}_{2\perp}) \equiv \int d^4x\, \Delta_1(x)\Delta_2(x)|_{\theta=\pi/2},
\end{equation}
where the subscript 
means that the integral is evaluated
setting $\theta=\pi/2$ in the quantities~(\ref{eq:ILMthetaquant}) and ~(\ref{eq:ILMthetaquant2}).
A comparison with the result of Ref.~\cite{instanton1} is in order. In that
paper the authors perform the calculation of the correlation function
only, but they do not take into account the ef\mbox{}fect of the pseudoparticles
interacting with the two loops separately, which is canceled only after
dividing the correlation function by the expectation values of the
loops: their result [Eq.~(63) of Ref.~\cite{instanton1}] is stated
in terms of an integral which is divergent, since the integrand is not
vanishing at inf\mbox{}inity, while in our calculation we f\mbox{}ind a f\mbox{}inite result. 
We have found the same analytic dependence on the angle $\theta$ as
in Ref.~\cite{instanton1}, but the prefactor
$n(2/N_c)^2F(\vec{z}_{\perp},\vec{R}_{1\perp},\vec{R}_{2\perp})$ in
Eq.~(\ref{eq:C_inst}) can now be assessed numerically. 

\begin{figure}[t]
  \centering
  \includegraphics[width=0.87\textwidth]{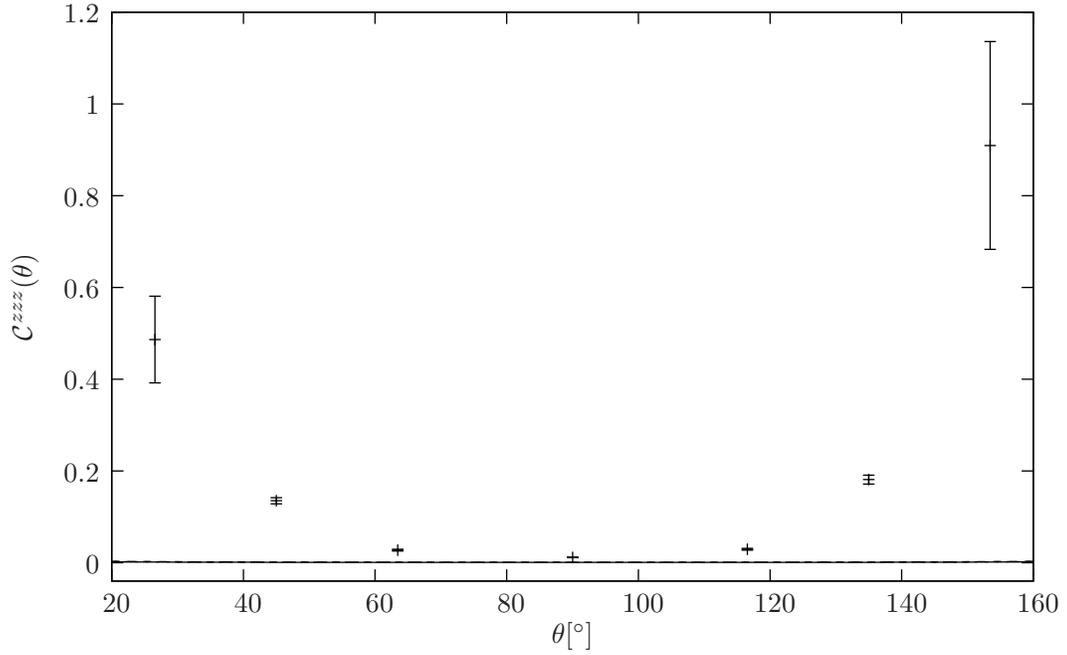}
  \caption[Comparison of the ILM prediction with the lattice
    data for the ``$zzz/zyy$'' conf\mbox{}iguration at $d=0$.]{Comparison of the ILM prediction, 
    Eqs.~(\ref{eq:C_inst})--(\ref{eq:inst_int_num}), with the lattice
    data for the ``$zzz/zyy$'' conf\mbox{}iguration at $\z=0\,\fm$. Here and in the
    following f\mbox{}igures the dotted line corresponds to $n_q=1.33\,{\rm fm}^{-4}$
    and the dashed line corresponds to $n_q=1.64\,{\rm   fm}^{-4}$, with
    $\rho_q=0.35\,{\rm fm}$; the solid line corresponds to the
    phenomenological values $n=1\,{\rm fm}^{-4}$ and $\rho=1/3\,{\rm fm}$.}
  \label{fig:instpred1}
\end{figure}

\begin{figure}
  \centering
  \includegraphics[width=0.9\textwidth]{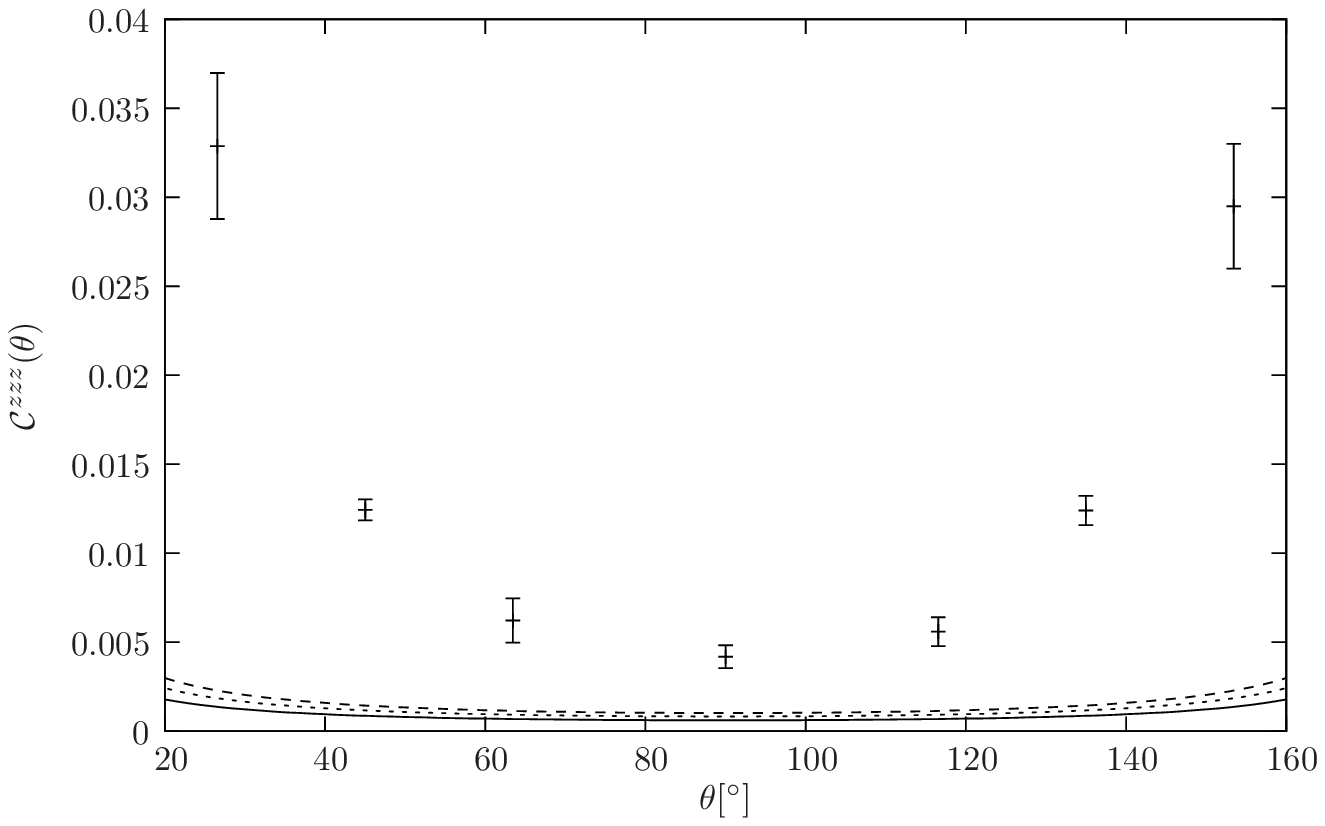}
  \caption{Same comparison as in Fig.~\ref{fig:instpred1} for the ``$zzz$'' conf\mbox{}iguration at $\z=0.1\,\fm$.}
  \label{fig:instpred2}
\end{figure}

\begin{figure}
  \centering
  \includegraphics[width=0.9\textwidth]{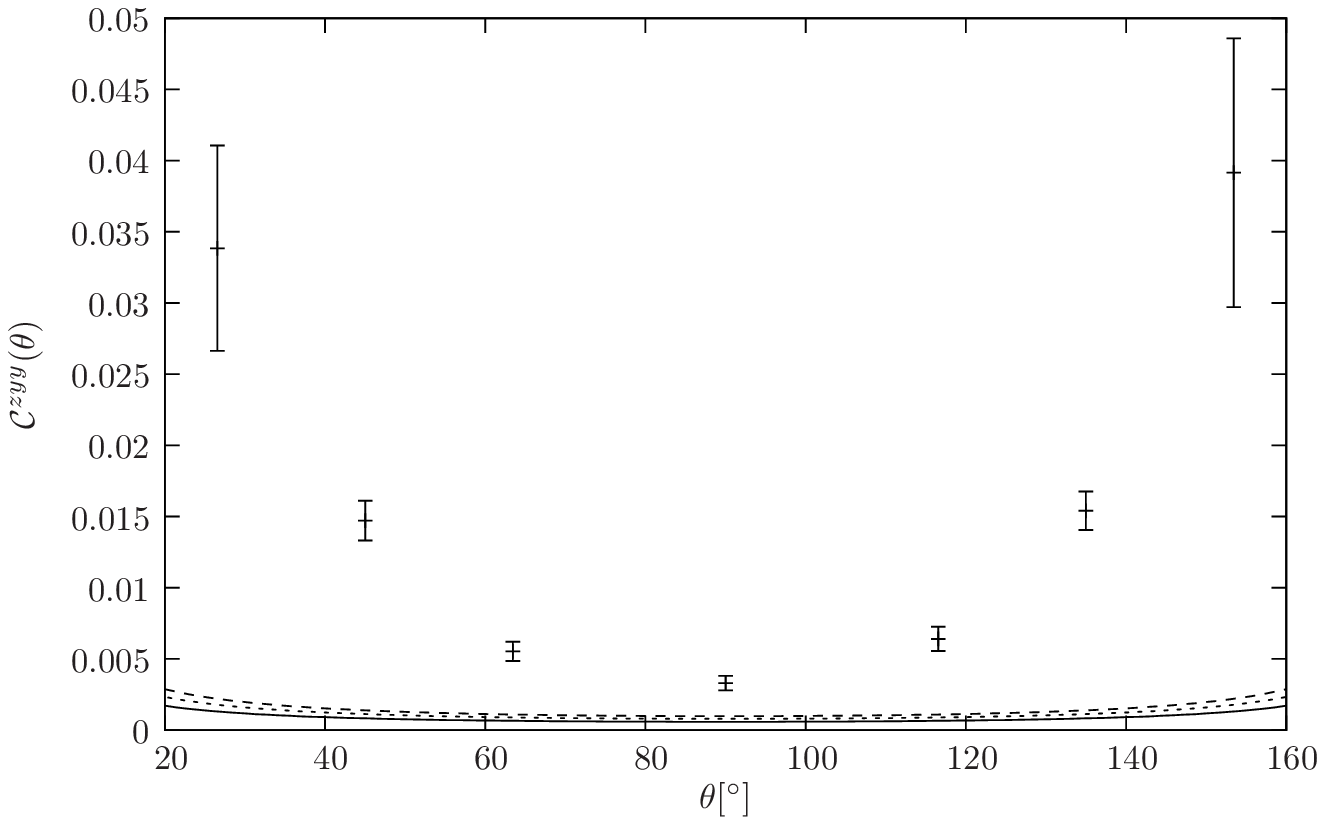}
  \caption{Same comparison as in Fig.~\ref{fig:instpred1} for the ``$zyy$'' conf\mbox{}iguration at $\z=0.1\,\fm$.}
  \label{fig:instpred3}
\end{figure}

\begin{figure}
  \centering
  \includegraphics[width=0.9\textwidth]{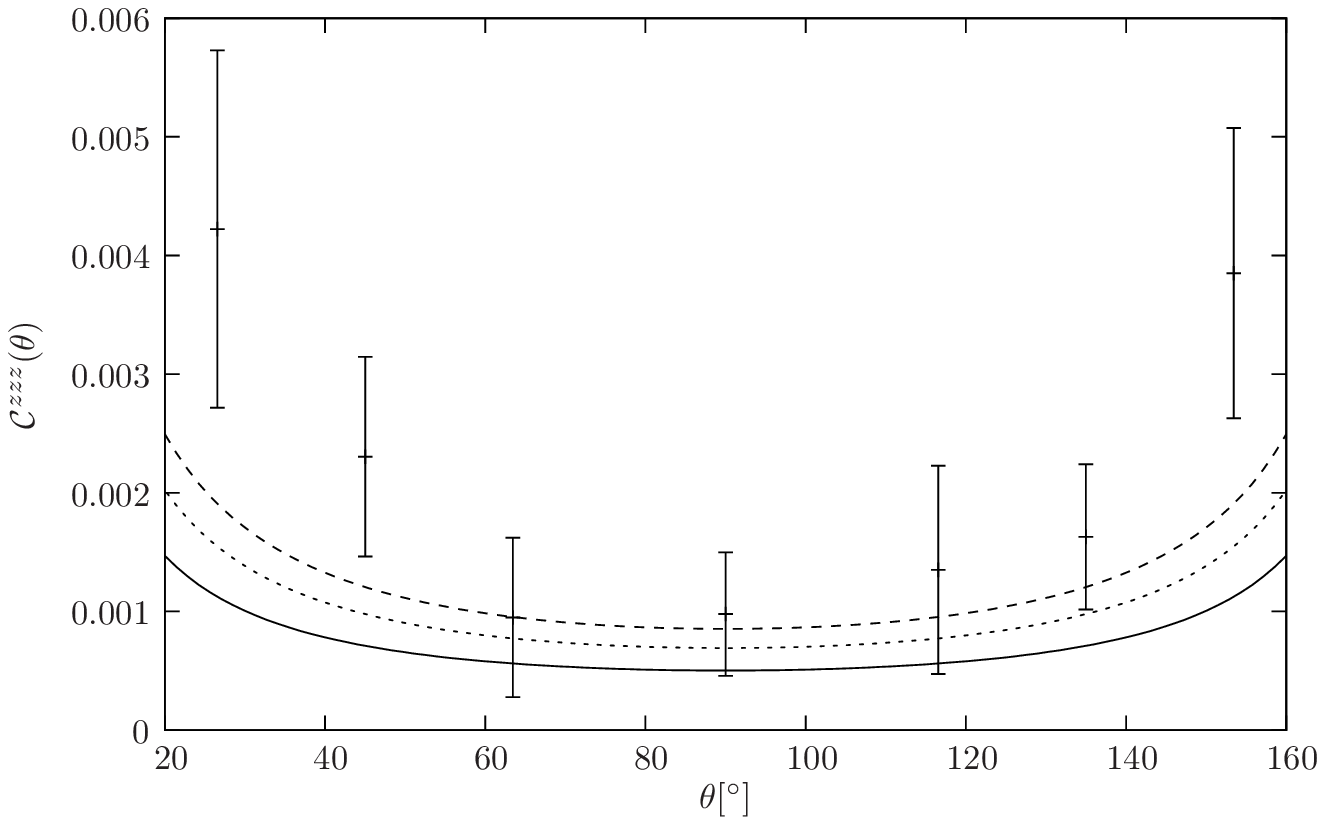}
  \caption{Same comparison as in Fig.~\ref{fig:instpred1} for the ``$zzz$'' conf\mbox{}iguration at $\z=0.2\,\fm$.}
  \label{fig:instpred4}
\end{figure}

\begin{figure}
  \centering
  \includegraphics[width=0.9\textwidth]{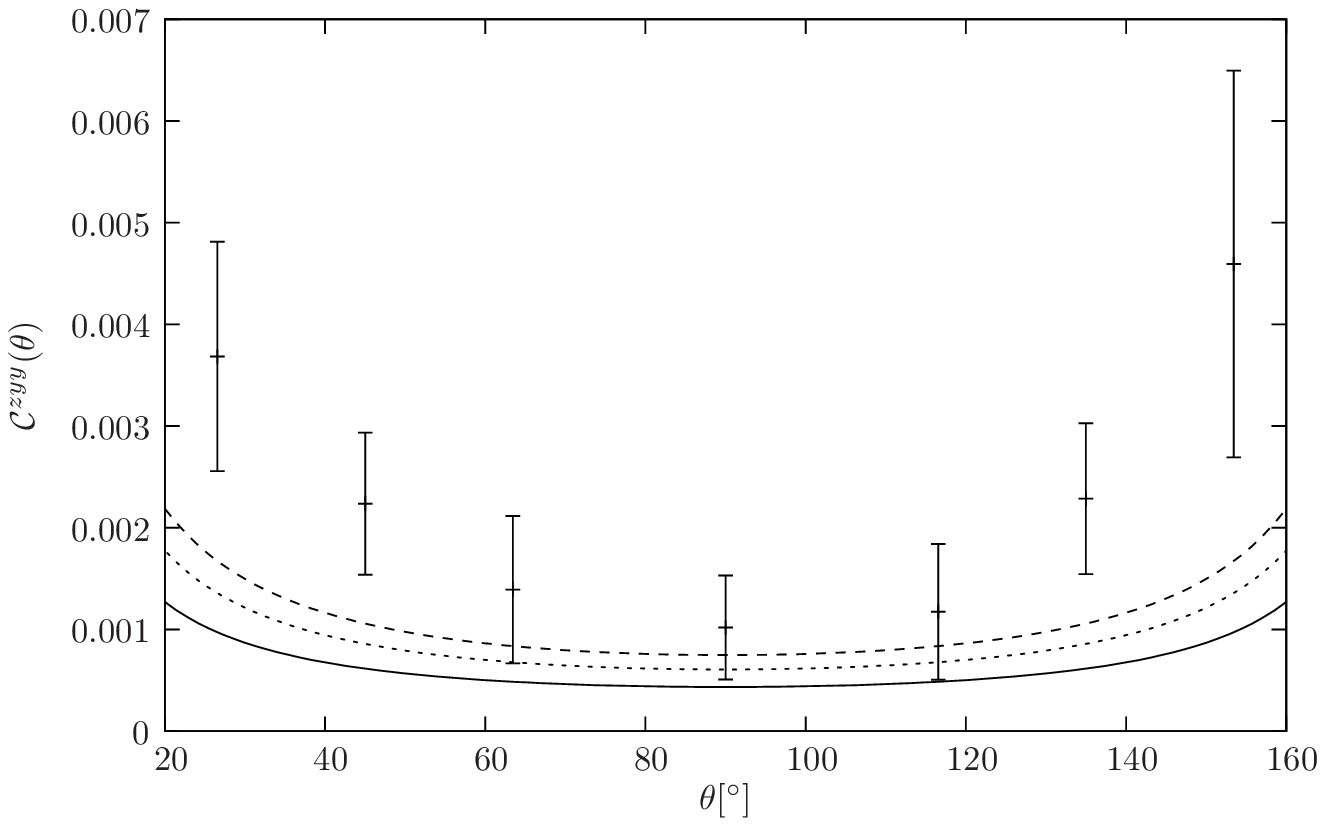}
  \caption{Same comparison as in Fig.~\ref{fig:instpred1} for the ``$zyy$'' conf\mbox{}iguration at $\z=0.2\,\fm$.}
  \label{fig:instpred5}
\end{figure}

In order to compare the analytic expression (\ref{eq:C_inst}) to the
lattice data, we have to set the density $n$ and the radius $\rho$ to
the appropriate values. The phenomenological estimates $n=1\,{\rm
  fm}^{-4}$ and $\rho=1/3\,{\rm fm}$ correspond to the {\it physical}
(i.e., {\it unquenched}) case, 
while our numerical simulations
have been performed in the {\it quenched} approximation of QCD, neglecting
dynamical--fermion ef\mbox{}fects. We
have then preferred to use the values $n_q=1.33$--$1.64\,{\rm
  fm}^{-4}$ for the density and  $\rho_q=0.35\,{\rm fm}$ for the
average size of the pseudoparticles, obtained directly by lattice
calculations in the pure--gauge theory~\cite{Chu}. 
The {\it packing fraction} is a factor $1.6$--$2$ 
larger than the one obtained with the phenomenological values, but it
is still a small number ($\lesssim 0.1$), so that the {\it dilute liquid}
picture is still reasonable\footnote{The phenomenological estimate of
  $n$ is obtained assuming that instantons and anti--instantons
  dominate the {\it gluon condensate}
  $G_2=\la(\alpha_s/\pi)F_{\mu\nu}^a
  F_{\mu\nu}^a\ra$~\cite{Shu,inst-rev}, so that $n\propto
  G_2$. If we assumed that the same holds in the pure--gauge 
  theory, we would obtain $n_q/n=G_2^q/G_2\simeq 5.8$ (using for $G_2$
  the value obtained in Ref.~\cite{Narison}, and for the {\it quenched
    gluon condensate} $G_2^q$ the result obtained on the lattice in
  Ref.~\cite{latticeSVM}), i.e., a value for the pseudoparticle
  density $n_q$ considerably larger than the one measured  on the
  lattice.}.  
The ILM prediction obtained with these 
values is shown in Figs.~\ref{fig:instpred1}--\ref{fig:instpred5},
together with the results obtained on the lattice in Ref.~\cite{lattice}, for
the loop conf\mbox{}igurations ``$zzz$'' ($\vec{R}_{1\perp}\para \vec{R}_{2\perp}\para
\vec{z}_{\perp}$) and ``$zyy$'' ($\vec{R}_{1\perp}\para \vec{R}_{2\perp}\perp
\vec{z}_{\perp}$), with $|\vec{R}_{1\perp}|=|\vec{R}_{2\perp}|=0.1\,{\rm
fm}$. The
dotted line corresponds to $n_q=1.33\,{\rm 
  fm}^{-4}$, while the dashed line corresponds to $n_q=1.64\,{\rm
  fm}^{-4}$; for comparison, we plot also the prediction obtained using the
phenomenological values of $n$ and $\rho$ (solid line).  As already
noticed in~\cite{lattice}, the 
functional form does not seem to be the correct one to properly describe the
lattice data, and a second term, proportional to $(\cot\theta)^2$, must
be added to obtain a good f\mbox{}it. In Table \ref{tab:insttab} we show
for comparison the ILM {\it prediction} of the prefactor 
$K=n_q({2}/{N_c})^2F(\vec{z}_{\perp},
\vec{R}_{1\perp},\vec{R}_{2\perp})$, 
and the value obtained with a f\mbox{}it to the lattice data with the f\mbox{}itting
functions (see Ref.~\cite{lattice}) 
\begin{equation}
  \label{eq:fitting}
  f_{\rm ILM} = K_{\rm ILM}\frac{1}{\sin\theta}, \qquad f_{\rm ILMp} = K_{\rm
  ILMp}\frac{1}{\sin\theta} + K_{\rm ILMp}' (\cot\theta)^2,
\end{equation}
where $f_{\rm ILMp}$ is obtained by adding the lowest--order perturbative
expression~\cite{BB,LLCM1,Meggiolaro05} to the ILM contribution. Notice,
however, that $K_{\rm ILMp}'$ can also receive nonperturbative contributions 
from two--instanton effects~\cite{instanton1}.

\begin{figure}[t]
  \centering
  \includegraphics[width=0.8\textwidth]{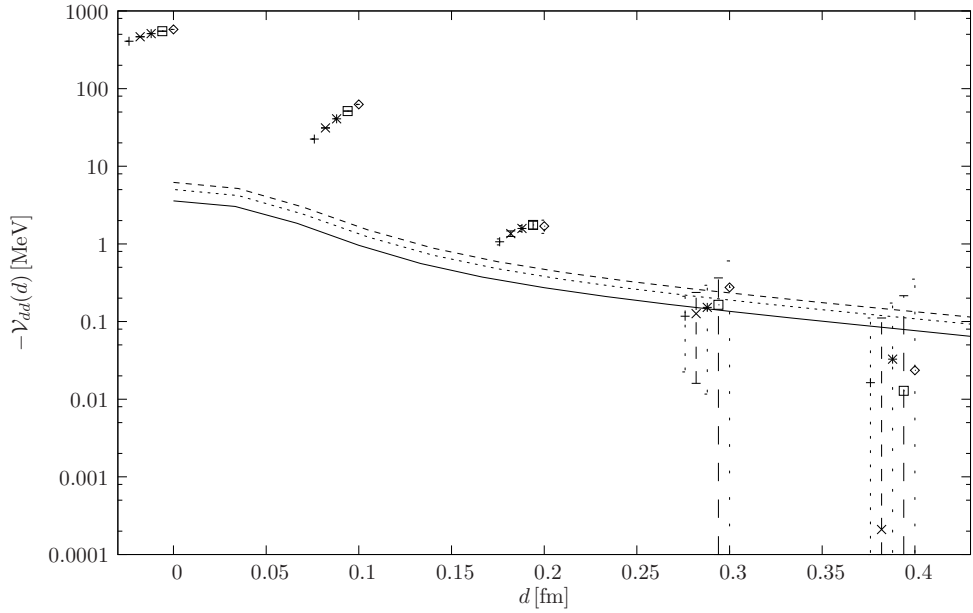}
  \caption[The instanton--induced dipole--dipole potential compared to the
    lattice data (on a logarithmic scale). Here $\vec{R}_1=\vec{R}_2$
    are parallel to $\vec{d}$, with $|\vec{R}_i|=0.1\,{\rm fm}$.]{The
    instanton--induced 
    dipole--dipole potential compared to the 
    lattice data (on a logarithmic scale). The three lines correspond to
    $n_q=1.33 \,{\rm fm}^{-4}$ (dotted line) and $n_q=1.64\,{\rm fm}^{-4}$
    (dashed line) with $\rho_q=0.35\,{\rm fm}$, and to the phenomenological
    values $n=1\,{\rm fm}^{-4}$ and $\rho =1/3\,{\rm fm}$ (solid
    line). Here $\vec{R}_1=\vec{R}_2$ are parallel to $\vec{d}$, with
    $|\vec{R}_i|=0.1\,{\rm fm}$. } 
\label{fig:ddpot_pp}
\end{figure}

\begin{figure}
  \centering
  \includegraphics[width=0.8\textwidth]{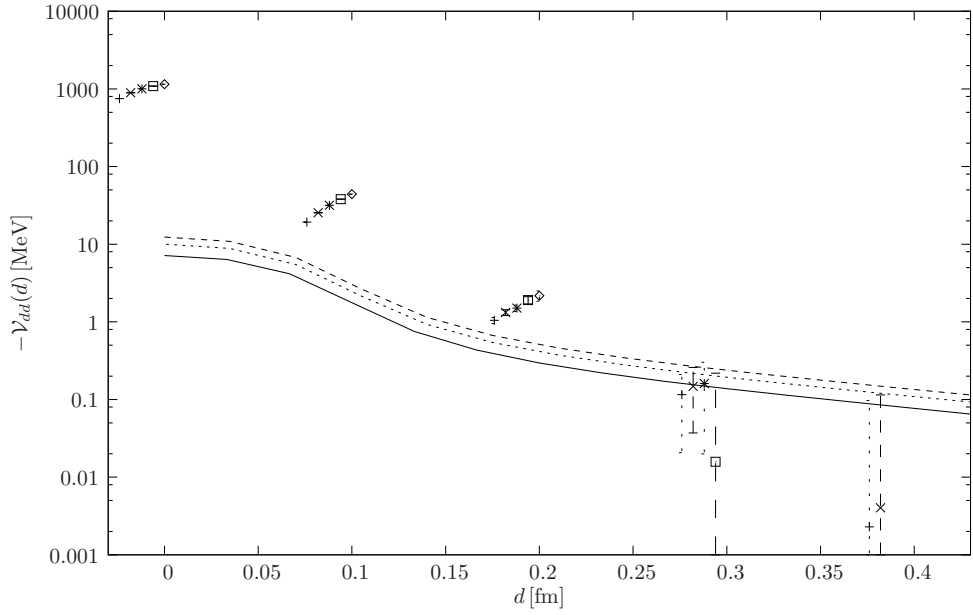}
  \caption{    Same comparison as in
    Fig.~\ref{fig:ddpot_pp}, but with $\vec{R}_1=-\vec{R}_2$,
     $\vec{R}_1$ parallel to $\vec{d}$, with
    $|\vec{R}_i|=0.1\,{\rm fm}$.} 
\label{fig:ddpot_app}
\end{figure}

\begin{figure}
  \centering
  \includegraphics[width=0.8\textwidth]{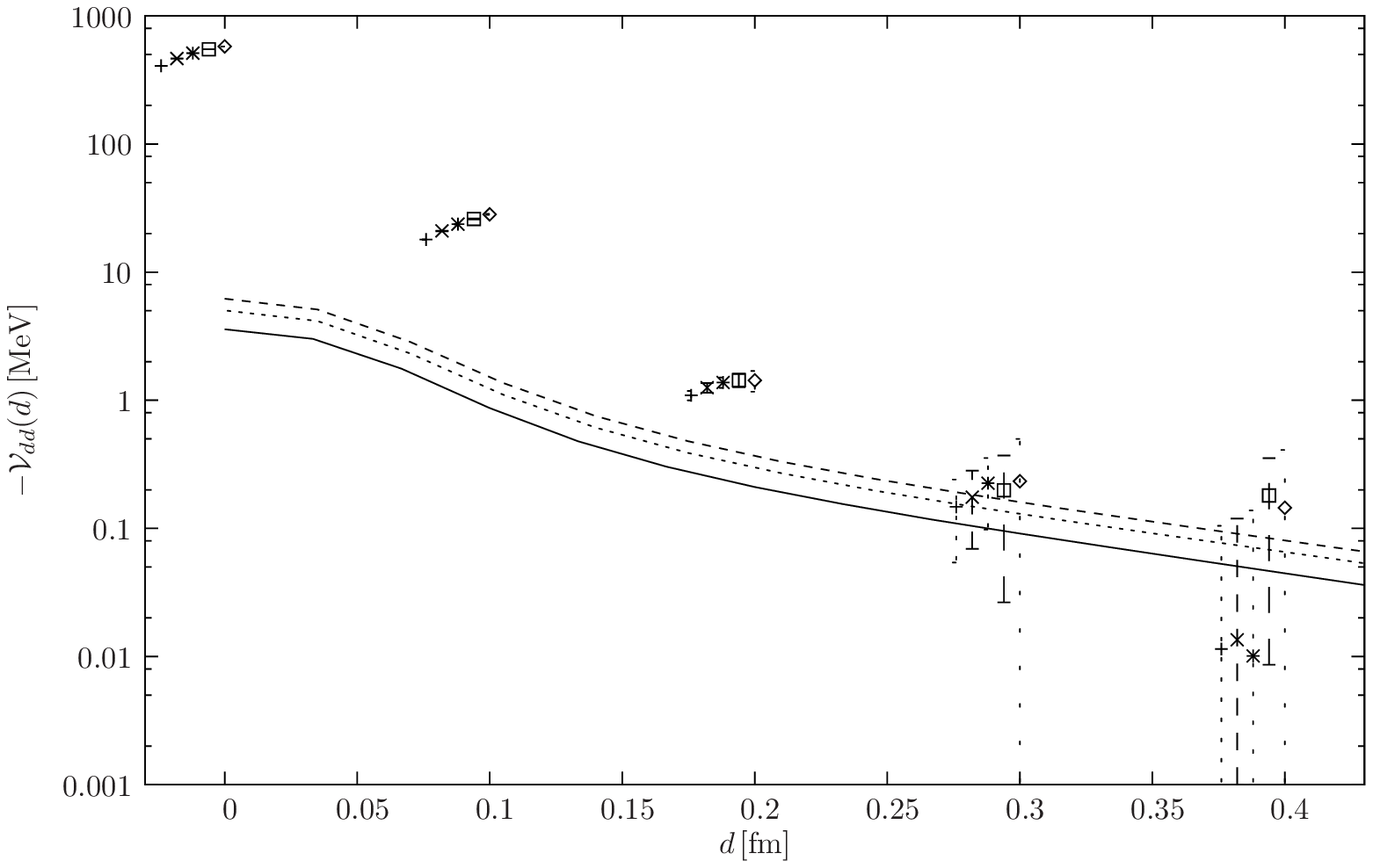}
  \caption{    Same comparison as in
    Fig.~\ref{fig:ddpot_pp}, but with $\vec{R}_1=\vec{R}_2$ orthogonal
    to $\vec{d}$, with $|\vec{R}_i|=0.1\,{\rm fm}$.}
\label{fig:ddpot_po}
\end{figure}

\begin{figure}
  \centering
  \includegraphics[width=0.8\textwidth]{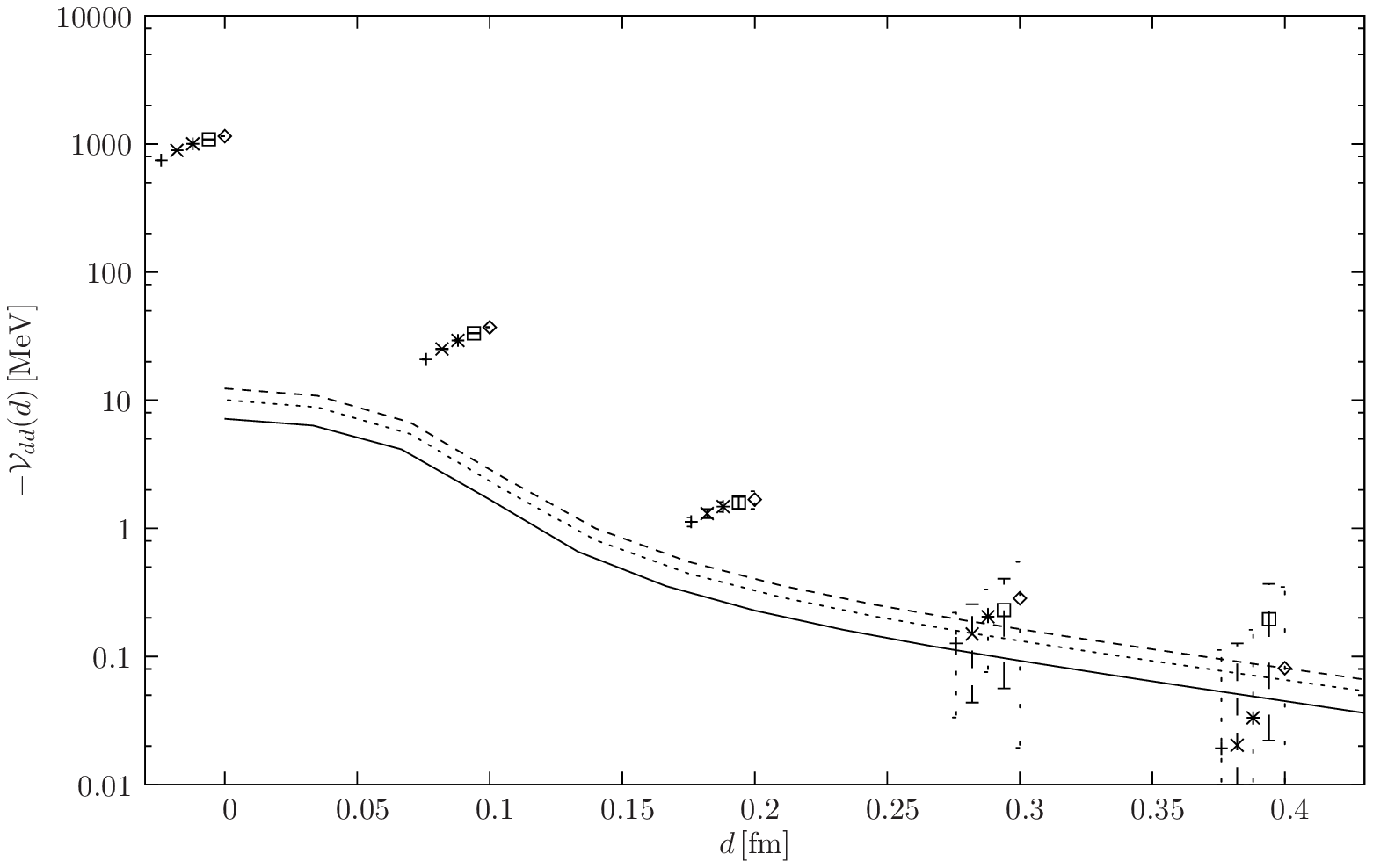}
  \caption{    Same comparison as in
    Fig.~\ref{fig:ddpot_pp}, but with $\vec{R}_1=-\vec{R}_2$
    orthogonal to $\vec{d}$, with
    $|\vec{R}_i|=0.1\,{\rm fm}$.}   
\label{fig:ddpot_apo}
\end{figure}

The instanton prediction turns out to be more or less of the correct
order of magnitude in the range of distances considered, at least around
$\theta=\pi/2$ [where, as we have said before, the one--instanton
  approximation used to derive the result, Eqs.~(\ref{eq:C_inst}) and
  (\ref{eq:inst_int_num}), is expected to make sense], but it does not
match the lattice data properly. The agreement with the data seems to
be quite good at $\z=0.2\,\fm$; however, concerning the dependence on
the relative distance between the loops, it seems that the ILM
overestimates the correlation length which sets the scale for the
rapid decrease of the correlation function. This is also supported by
the comparison of the instanton--induced {dipole--dipole potential},
which we calculate in the next Section, with some preliminary
numerical results on the lattice, as we show in
Figs.~\ref{fig:ddpot_pp}--\ref{fig:ddpot_apo}.

\newsection{Wilson--loop correlation function and the di\-pole--dipole
  potential}  
\label{sec:inst_dd}

In this Section we calculate the normalised correlation function of
two (inf\mbox{}inite) parallel Wilson loops, which describe the time
evolution of two static colour dipoles.
The one--instanton approximation makes no sense here, since the
interaction region has inf\mbox{}inite extension in the time direction, and
the ef\mbox{}fect of a whole time--series of pseudoparticles on the
two dipoles has to be 
considered. 
The above--mentioned correlation function can be used to extract the
{dipole--dipole potential} ${\cal V}_{dd}$ 
by means of the formula~\cite{Pot} 
\begin{equation}
  \label{eq:ddpot}
  \dfrac{ \langle {\cal W}_1 {\cal W}_2
    \rangle} {\langle {\cal W}_1 \rangle \langle
    {\cal W}_2 \rangle } \mathop\simeq_{T\to\infty} e^{-T{\cal
      V}_{dd}(\vec{d},\vec{R}_1,\vec{R}_2)}, 
\end{equation}
where $\vec{R}_1$ and $\vec{R}_2$ are the sizes of the two dipoles,
$\vec{d}$ is the distance between their centers, and $T$ is the length of
the two loops. Neglecting again the transverse connectors, the loops
${\cal W}_{1,2}$ are written as
\begin{equation}
  {\cal W}_i = \frac{1}{N_c}\tr
  [{\rm W}^{(-)\dag}_i{\rm W}^{(+)}_i] ,
\end{equation}
with
\begin{equation}
\begin{aligned}
  {\rm W}^{(\pm)}_1 &= {\cal P} \exp\left[ -ig \displaystyle\int_{0}^{T}
      A_{4}(ut + d \pm R_1/2) \, dt \right],\\
  {\rm W}^{(\pm)}_2 &= {\cal P} \exp\left[ -ig \displaystyle\int_{0}^{T}
      A_{4}(ut\pm R_2/2) \, dt \right],
\end{aligned}
\end{equation}
where $u = (\vec{0},1)$, $R_i = (\vec{R}_i,0)$, $d = (\vec{d},0)$. To
evaluate their correlation function [the numerator in (\ref{eq:ddpot})]
we divide space--time into time--slices, and, 
in turn, the time--slices into cells; the procedure is the same as for the
expectation value of a single loop, but this time we determine the
spatial size $V^{(3)}_{\rm cell}$ to be large enough to accommodate
both the dipoles. 
Requiring
\begin{equation}
   V_{\rm cell}^{(3)} \sim \left(|\vec{d}| + \frac{|\vec{R}_1|+|\vec{R}_2|}{2}
   + 2\rho\right) \left(\frac{|\vec{R}_1|+|\vec{R}_2|}{2} + 2\rho\right)^2,
\end{equation}
we have to set $\delta T$ to
\begin{equation}
\begin{aligned}
   \frac{\delta T}{\rho} &\sim 4, &{\rm for }\,
   |\vec{d}|,|\vec{R}_i|\sim \rho/2, ~~~~~
   \frac{\delta T}{\rho} &\sim 2.5, &{\rm for }\,
   |\vec{d}|,|\vec{R}_i|\sim \rho,
\end{aligned}
\end{equation}
which is still a fairly large value. Note however that we are
overestimating the volume of the region of 3--space where a
pseudoparticle can af\mbox{}fect the loop, so that $\delta T$ is actually
underestimated. Again, for large distances the two dipoles interact
with dif\mbox{}ferent pseudoparticles, and the correlation function is
expected to factorise, thus giving a vanishing dipole--dipole potential
as $|\vec{d}|\to \infty$. 

\begin{figure}[t]
  \centering
  \includegraphics[width=0.7\textwidth]{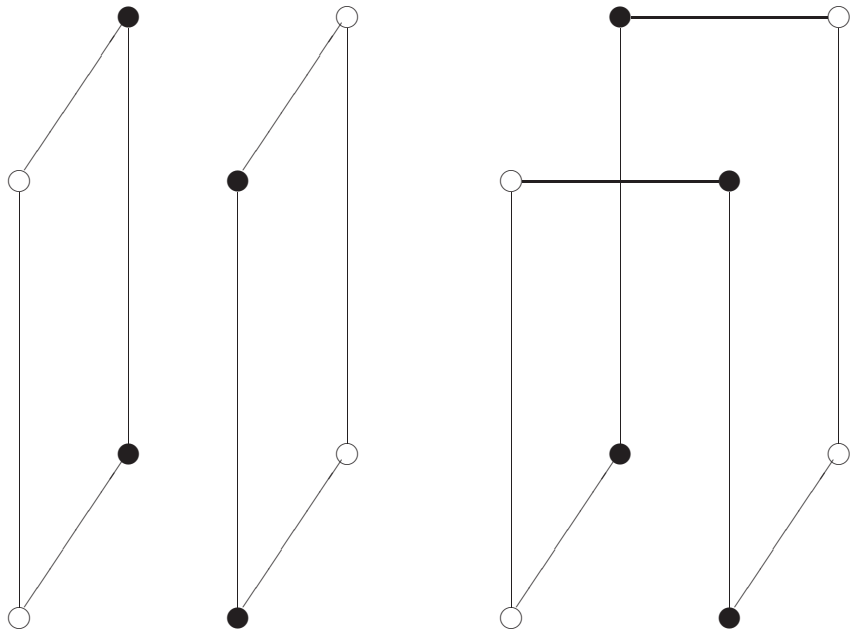}
  \caption[Schematic representation of the operators ${\cal W}_1{\cal
      W}_2$ and $\overline{\cal W}_{12}$.]{Schematic representation of
    the operators ${\cal W}_1{\cal 
      W}_2$ (left) and $\overline{\cal W}_{12}$ (right), corresponding
    respectively to the processes $dd\to dd$ and $dd\to
    \widetilde{d}\widetilde{d}$. White (black) circles represent
    quarks (antiquarks).}  
\label{fig:operators}
\end{figure}

Now let $K$ be the average number of instantons that interact with the
loop. This number grows linearly with the length $T$ of the loop due
to the uniformity of the liquid, and we can take, without loss of
generality, $T=K\delta T$, since we are interested in the limit
$T\to\infty$. We divide space--time into cells as described above,
labelling $D_1,\ldots,D_K$ the cells where the loop lives, and we def\mbox{}ine
\begin{equation}
\begin{aligned}
  {\rm W}^{(\pm)}_1(k) &= {\cal P} \exp\left[ -ig \displaystyle\int_{(k-1)\delta
      T}^{k\delta T}  A_{4}(ut + d \pm R_1/2) \, dt \right],\\
  {\rm W}^{(\pm)}_2(k) &= {\cal P} \exp\left[ -ig \displaystyle\int_{(k-1)\delta
      T}^{k\delta T}  A_{4}(ut\pm R_2/2) \, dt \right],
  \quad k=1,\ldots,K,
\end{aligned}
\end{equation}
and the ``two--link'' variables
\begin{equation}
\begin{aligned}
  \T_i(k) &= {\rm W}^{(-)*}_i(k)\otimes {\rm W}^{(+)}_i(k),\qquad
  \W_i(j) &= \prod_{k=j}^{1}\T_i(k),
\end{aligned}
\end{equation}
which allow us to write
\begin{equation}
    {\cal W}_i = \frac{1}{N_c} \left[\W_i(K)\right]_{ijij} =
    \frac{1}{N_c}\ttr \W_i(K).
\end{equation}
The integration over the colour degrees of freedom of the
pseudoparticles is more complicated than in the case of a single loop,
although the procedure is the same. 
Starting from the $K$--th pseudoparticle, we have to calculate the
integral 
\begin{align}
&\begin{aligned}
   {\cal I}_1 &\equiv \int dU(K) \ttr\W_1(K) \ttr\W_2(K)   \\ &=\int dU(K)
   \hat{\T}_1(K)_{j_1k_1}\check{\W}_1(K-1)_{j_1k_1}\hat{\T}_2(K)_{j_2k_2}\check{\W}_2(K-1)_{j_2k_2}\\
   &=I_{j'_1k'_1j_1k_1j'_2k'_2j_2k_2}\hat{\T}_1(K)_{j'_1k'_1}|_s\check{\W}_1(K-1)_{j_1k_1}
\hat{\T}_2(K)_{j'_2k'_2}|_s\check{\W}_2(K-1)_{j_2k_2},
\end{aligned}
\intertext{where the subscript $s$ means
that the given quantity has to be evaluated in the f\mbox{}ield of a
pseudoparticle with standard colour orientation, where
we have used the notation}
  &\hat{\T}_i(k)_{jk} = \T_i(k)_{ijik}, ~~  \check{\W}_i(k)_{jk} =
  \W_i(k)_{jiki}, ~~ \ttr\W_i(k) = \hat{\T}_i(k)_{jk}\check{\W}_i(k-1)_{jk},
\intertext{and where~\cite{Creutz}} 
&\begin{aligned}
&    I_{j'_1k'_1j_1k_1j'_2k'_2j_2k_2} = \int dU\,
    U_{j_1j'_1}U^\dag_{k'_1k_1}U_{j_2j'_2}U^\dag_{k'_2k_2}\\=&
    \left(a\delta_{j'_1k'_1}\delta_{k'_2k'_2} +
    b\delta_{j'_1k'_2}\delta_{j'_2k'_1}\right)\delta_{j_1k_1}\delta_{j_2k_2} +
    \left(b\delta_{j'_1k'_1}\delta_{k'_2k'_2} +
    a\delta_{j'_1k'_2}\delta_{j'_2k'_1}\right)\delta_{j_1k_2}\delta_{j_2k_1},\\
   &a = \frac{1}{N_c^2-1}, \quad b = -\frac{1}{N_c}a .
\end{aligned}
\intertext{Performing the contractions of colour indices we obtain}
&\begin{aligned}
 {\cal I}_1 &=   a\big[N_c^2w_1(K)w_2(K) -\underline{w}_{12}(K)\big]
\ttr{\W}_1(K-1)\ttr{\W}_2(K-1)  \\
& \phantom{unbelpezzounbelpezzounb}  +aN_c\big[\underline{w}_{12}(K)- w_1(K)w_2(K)\big]
\overline{\mathbb{W}}_{12}(K-1),
\end{aligned}
\end{align}
where\footnote{Here and in the following we adopt the following short--hand
notation for quantities which depend on the position and the type of
the pseudoparticle in the $k$--th cell,
$f(k)=f(z_{P(k)},\sigma_{P(k)})$.}
\begin{align}
&\begin{aligned}
  w_i(k) &\equiv \frac{1}{N_c}\hat{\T}_i(k)_{j'_1j'_1}|_s = \frac{1}{N_c}\tr
  [{\rm W}^{(-)\dag}_i(k){\rm W}^{(+)}_i(k)]|_s =
  1-\frac{2}{N_c}\Delta_i(z_{P(k)}),\\ 
  \Delta_i(z) &= 1-\cos\alpha_{i+}\cos\alpha_{i-} -
  \hat{n}_{i+}\cdot\hat{n}_{i-}\sin\alpha_{i+}\sin\alpha_{i-},\\
  \alpha_{i\pm} &= \frac{\pi\|n_{i\pm}\|}{\sqrt{\|n_{i\pm}\|^2 +
      \rho^2}},\quad
  n^a_{1\pm} = -(d \pm \frac{R_1}{2} -z)_a, \quad n^a_{2\pm} = -(\pm
  \frac{R_2}{2} -z)_a,
\end{aligned}
\intertext{and we have introduced the quantities}
  &\underline{w}_{12}(k)\equiv\frac{1}{N_c}\hat{\T}_1(k)_{j'_1j'_2}|_s\hat{\T}_2(k)_{j'_2j'_1}|_s
  = \frac{1}{N_c}\tr
  [{\rm W}^{(-)\dag}_1(k){\rm W}^{(+)}_1(k){\rm W}^{(-)\dag}_2(k){\rm
    W}^{(+)}_2(k)]|_s,  \nonumber\\ 
& \overline{\mathbb{W}}_{12}(k) \equiv \check{\W}_1(k)_{j_1j_2}\check{\W}_2(k)_{j_2j_1}.
\end{align}
{The effect of a single instanton on ${\rm W}^{(\pm)}_i(k)$ is given by}
\begin{equation}
  \label{eq:W_i2}
   {\rm W}_i^{(\pm)} 
  = \left( 
  \begin{array}{cc}
    \hspace{1mm} \exp \left[ i \hat{n}^{a}_{i\pm} \sigma^a \alpha_{i\pm}
      \right]              \hspace{1mm} & 
    \hspace{1mm} 0         \hspace{1mm} \vspace{1mm}\\
    \hspace{1mm} 0         \hspace{1mm}  & 
    \hspace{1mm} \I_{N_c-2} \hspace{1mm}
  \end{array}\right) ,
\end{equation} 
{where $\I_M$ is the $M$-dimensional unit matrix, while the effect of an
  anti--instanton is obtained by replacing 
$n_{i\pm}\to -n_{i\pm}$}. 
The expectation value of the quantity
\begin{equation}
  \overline{\cal W}_{12} \equiv
  \frac{1}{N_c}\overline{\mathbb{W}}_{12}(K) =
\frac{1}{N_c}\tr [{\rm W}^{(-)\dag}_2{\rm W}^{(+)}_1
  {\rm W}^{(-)\dag}_1 {\rm W}^{(+)}_2]
\end{equation}
can be interpreted as the transition amplitude of an {\it inelastic process}, the
f\mbox{}inal state $\widetilde{d}\widetilde{d}$ being the initial one, $dd$, with the (say)
antiquarks in the two dipoles interchanged (see Fig.~\ref{fig:operators}). Over the time $T$ a
certain number of such transitions can happen, and this is at the origin
of the new terms in the formulas above.

To proceed we also need to calculate the integral
\begin{equation}
    {\cal I}_2 \equiv \int dU(K)\overline{\mathbb{W}}_{12}(K).
\end{equation}
If we write
\begin{equation}
\begin{aligned}
&
  \overline{\mathbb{W}}_{12}(K)=\check{\W}_1(K)_{j_1j_2}\check{\W}_2(K)_{j_2j_1}=
           [\T_1(K)]_{j_1j'_1j_2k'_1}|_s[\T_2(K)]_{j_2j'_2j_1k'_2}|_s \\
& \times U_{j_1j'_1}U^\dag_{k'_1k_1}U_{j_2j'_2}U^\dag_{k'_2k_2}
\check{\W}_1(K-1)_{j_1k_1}\check{\W}_2(K-1)_{j_2k_2},
\end{aligned}
\end{equation}
we f\mbox{}ind the same Haar integral as before, and thus, after contracting
the colour indices, we f\mbox{}ind
\begin{equation}
\begin{aligned}
& {\cal I}_2 =  aN_c\big[\overline{w}_{12}(K)-
  \widetilde{w}_1(K)\widetilde{w}_2(K)\big] 
\ttr{\W}_1(K-1)\ttr{\W}_2(K-1) \\&\phantom{unbelpezzounbelpezzo} +
a\big[N_c^2 \widetilde{w}_1(K)\widetilde{w}_2(K)-\overline{w}_{12}(K)\big]
\overline{\W}_{12}(K-1),
\end{aligned}
\end{equation}
where we have introduced the quantity $\overline{w}_{12}(k)$,
\begin{equation}
\begin{aligned}
 & \overline{w}_{12}(k)\equiv
  \frac{1}{N_c}[\T_1(k)]_{j_1j'_1j_2j'_1}|_s[\T_2(k)]_{j_2j'_2j_1j'_2}|_s\\
 & \phantom{unbelpezzounbelpezzo}=     \frac{1}{N_c}\tr[  {\rm W}^{(+)}_1(k) {\rm W}^{(-)\dag}_1(k)
  {\rm W}^{(+)}_2(k)
  {\rm W}^{(-)\dag}_2(k)]|_s, 
\end{aligned}
\end{equation}
{and the quantities $\widetilde{w}_i(k)$, which come from the
  contraction of}
\begin{equation}
 \widetilde{w}_1(k) \widetilde{w}_2(k) =
 \frac{1}{N_c^2}[\T_1(k)]_{j_1j'_1j_2j'_2}|_s[\T_2(k)]_{j_2j'_2j_1j'_1}|_s, 
\end{equation}
{and are equal to the value of the Wilson loops (of
  inf\mbox{}inite length) obtained from ${\cal W}_i$ interchanging the
  antiquarks in the two dipoles, in the f\mbox{}ield of a single pseudoparticle:}
\begin{equation}
\begin{aligned}
 \widetilde{w}_1(k) &\equiv \frac{1}{N_c}\tr [{\rm W}^{(-)\dag}_2(k)
  {\rm W}^{(+)}_1(k)]|_s =1-\frac{2}{N_c}\widetilde{\Delta}_1(z_{P(k)}),\\  
 \widetilde{\Delta}_1(z) &= 1-\cos\alpha_{1+}\cos\alpha_{2-} -
  \hat{n}_{1+}\cdot\hat{n}_{2-}\sin\alpha_{1+}\sin\alpha_{2-},\\
\widetilde{w}_2(k)&\equiv \frac{1}{N_c}\tr[ {\rm W}^{(-)\dag}_1(k)
  {\rm W}^{(+)}_2(k)]|_s= 1-\frac{2}{N_c}\widetilde{\Delta}_2(z_{P(k)}),
  \\  
  \widetilde{\Delta}_2(z) &= 1-\cos\alpha_{2+}\cos\alpha_{1-} -
  \hat{n}_{2+}\cdot\hat{n}_{1-}\sin\alpha_{2+}\sin\alpha_{1-},
\end{aligned}
\end{equation}
where the functions $\widetilde{\Delta}_i$ are again independent of
the pseudoparticle species. Also, $\overline{w}_{12}(k)$ is related to
$\underline{w}_{12}(k)$ in the same way.

To iterate the process we organise the previous results as follows:
\begin{equation}
\begin{aligned}
  \int dU(K)&\left(
  \begin{array}{c}
    \ttr{\W}_1(K)\ttr{\W}_2(K) \vspace{1mm}\\
    \overline{\W}_{12}(K)
  \end{array}
\right) \\ & \phantom{unbelpezzo} =  M(K)
  \left(
  \begin{array}{c}
    \ttr{\W}_1(K-1)\ttr{\W}_2(K-1) \vspace{1mm} \\
    \overline{\W}_{12}(K-1)
  \end{array}
\right),
\end{aligned}
\end{equation}
where $M(k)$ is the matrix
\begin{equation}
  M(k) =   \frac{1}{N_c^2-1}\left(
  \begin{array}{cc}
N_c^2w_1(k)w_2(k) -\underline{w}_{12}(k) &
N_c\big[\underline{w}_{12}(k) -  w_1(k)w_2(k)\big] \vspace{1.5mm}  \\ 
N_c\big[\overline{w}_{12}(k)-\widetilde{w}_1(k)\widetilde{w}_2(k)\big] &
N_c^2 \widetilde{w}_1(k)\widetilde{w}_2(k) - \overline{w}_{12}(k)
  \end{array}
\right).
\end{equation}
The iteration is now straightforward, and yields
\begin{equation}
\begin{aligned}
  \int d^KU &
  \left(
  \begin{array}{c}
    N_c^2 {\cal W}_1 {\cal W}_2\vspace{1mm}\\
    N_c \overline{\cal W}_{12}
  \end{array}
\right) \\=
  & \int d^KU 
  \left(
  \begin{array}{c}
    \ttr{\W}_1(K)\ttr{\W}_2(K) \vspace{1mm}\\
    \overline{\W}_{12}(K)
  \end{array}
\right) = \left[\prod_{k=K}^{1}M(k)\right]
  \left(
  \begin{array}{c}
    N_c^2 \vspace{1.5mm}\\
    N_c
  \end{array}
\right),
\end{aligned}
\end{equation}
where we used the fact that
\begin{align}
  \left(
  \begin{array}{c}
    \ttr{\W}_1(1)\ttr{\W}_2(1) \vspace{1mm}\\
    \overline{\W}_{12}(1)
  \end{array}
\right)\bigg|_s =
  \left(
  \begin{array}{c}
    {N_c^2}w_1(1)w_2(1) \vspace{1.5mm}\\
    {N_c}\overline{w}_{12}(1)
  \end{array}
\right)= M(1)  \left(
  \begin{array}{c}
    N_c^2 \vspace{1.5mm}\\
    N_c
  \end{array}
\right).
\end{align}
We are left with the integration over the pseudoparticle
positions. The procedure is described in Appendix \ref{app:C}; 
using Eq.~(\ref{eq:inst-av-over-pos-2}) we then obtain 
\begin{equation}
\begin{aligned}
\int d^{4N}z \, {\cal D}(\{z\}) \left[\prod_{k=K}^{1}M(k)\right] =
   \prod_{j=K}^{1} n\int_{D_j} d^{4}z \, 
   \Big(\nu_{\inst}M(z,\sigma_{\inst}) +
  \nu_{\ainst}M(z,\sigma_{\ainst})\Big)\\ =
\prod_{j=K}^{1} \left[\I_2 + n\int_{D_j} d^{4}z \,
   \Big(\nu_{\inst} \hat{M}(z,\sigma_{\inst}) +
  \nu_{\ainst} \hat{M}(z,\sigma_{\ainst})\Big)\right],
\end{aligned}
\end{equation}
where $\hat{M}\equiv M-\I_2$ falls of\mbox{}f to zero as
$|z|\to\infty$, and where $\nu_{\inst,\ainst}\equiv n_{\inst,\ainst}/n$. 
Performing the trivial integration over the time position, extending
 the spatial integration to the whole space, and letting (formally) $K\to
\infty$ with $T$ f\mbox{}ixed, we obtain
\begin{align}
&\int d^{4N}z \,{\cal D}(\{z\}) \left[\prod_{k=K}^{1}M(k)\right] =  \exp\{nT{\cal J} \},
\end{align}
where the matrix ${\cal J}$ is given by
\begin{align}
  {\cal J} &  \equiv \int d^{3}z \Big(\nu_{\inst} \hat{M}(z,\sigma_{\inst}) +
   \nu_{\ainst} \hat{M}(z,\sigma_{\ainst})\Big)
  =  \frac{1}{N_c^2-1}\left(
  \begin{array}{cc}
    N_c^2A -B & N_c(B-A)\vspace{1mm}\\
    N_c(\widetilde{B}-\widetilde{A}) &     N_c^2\widetilde{A} -\widetilde{B}
  \end{array}\right)\nonumber\\
&= \left(
  \begin{array}{cc}
    B & 0 \vspace{1mm}\\
    0 & \widetilde{B}
  \end{array}\right) + \frac{1}{N_c^2-1}\left(
  \begin{array}{cc}
    N_c^2(A -B) & N_c(B-A) \vspace{1mm}\\
    N_c(\widetilde{B}-\widetilde{A}) &     N_c^2(\widetilde{A} -\widetilde{B})
  \end{array}\right) ,
\end{align}
having def\mbox{}ined
\begin{equation}
\label{eq:ddpotfunctions}
\begin{aligned}
 A  &=\int d^{3}z \left[\left(\frac{2}{N_c}\right)^2\Delta_1(z)\Delta_2(z) -
    \frac{2}{N_c}\Delta_1(z) - \frac{2}{N_c}\Delta_2(z)\right],\\
  \widetilde{A}  &=\int d^{3}z \left[\left(\frac{2}{N_c}\right)^2\widetilde{\Delta}_1(z)\widetilde{\Delta}_2(z) -
    \frac{2}{N_c}\widetilde{\Delta}_1(z) - \frac{2}{N_c}\widetilde{\Delta}_2(z)\right],\\
  B &= \int d^{3}z \,\bigg[\nu_{\inst} (\underline{w}_{12}(z,\inst)-1) + 
  \nu_{\ainst}(\underline{w}_{12}(z,\ainst)-1)\bigg],\\
  \widetilde{B} &= \int d^{3}z\, \bigg[\nu_{\inst} (\overline{w}_{12}(z,\inst)-1) +
  \nu_{\ainst}(\overline{w}_{12}(z,\ainst)-1)\bigg].
\end{aligned}
\end{equation}
To proceed further with the calculation we now set the pseudoparticle
fractions to their phenomenological values $\nu_{\inst}=\nu_{\ainst} =
1/2$: in this case we can show that 
 (see Appendix \ref{app:C})
\begin{equation}
\begin{aligned}
    B = \widetilde{B} &= \frac{2}{N_c}\int d^{3}z \bigg[\big(\Delta_1(z)\Delta_2(z)
    +\widetilde{\Delta}_1(z)\widetilde{\Delta}_2(z)-\Delta_+(z)\Delta_-(z)\big)
      \\ &\phantom{unbe}-
   \big(\Delta_1(z) +\Delta_2(z) +\widetilde{\Delta}_1(z) +\widetilde{\Delta}_2(z)
  -\Delta_+(z) - \Delta_-(z)\big)\bigg],
\end{aligned}
\end{equation}
{where the quantities}
\begin{equation}
\label{eq:deltapm}
\begin{aligned}
 {\Delta}_+(z) &= 1-\cos\alpha_{1+}\cos\alpha_{2+} -
  \hat{n}_{1+}\cdot\hat{n}_{2+}\sin\alpha_{1+}\sin\alpha_{2+},\\
  {\Delta}_-(z) &= 1-\cos\alpha_{2-}\cos\alpha_{1-} -
  \hat{n}_{2-}\cdot\hat{n}_{1-}\sin\alpha_{2-}\sin\alpha_{1-},
\end{aligned}
\end{equation}
are the {analogues} of $\widetilde{\Delta}_i$ with the positions of the quark
and the antiquark in the (say) second dipole interchanged.
It is now easy to diagonalise ${\cal J}$: denoting with $X=A-B$, 
$\widetilde{X}=\widetilde{A}-B$, the eigenvalues can be written as
\begin{equation}
  {\lambda}_{\pm} = \frac{A+\widetilde{A}}{2} +
  \frac{1}{2(N_c^2-1)}\left\{X+\widetilde{X} \pm N_c \sqrt{(N_c^2-1)(\widetilde{X}-X)^2+(\widetilde{X}+X)^2}\right\};
\end{equation}
since they are dif\mbox{}ferent\footnote{The only exception is the case
  $\widetilde{X}=X=0$, in which case however the matrix is already in 
diagonal form.}, ${\cal J}$ is diagonalisable. Denoting by $\Pi_\pm$
the corresponding projectors we can write [$v_0\equiv(1,0)$,
  $v_{N_c}\equiv (N_c^2,N_c)$]
\begin{equation}
\begin{aligned}
  N_c^2\la {\cal W}_1 {\cal W}_2 \ra_{\rm ILM} &= v_0 \cdot \big( \exp\{nT{\cal J}\} v_{N_c}\big)
\\&= v_0\cdot \left[ \big(e^{nT\lambda_+}\Pi_+ + e^{nT\lambda_-}\Pi_-\big)v_{N_c}\right],
\end{aligned}
\end{equation}
and, taking the logarithm, in the large--$T$ limit we f\mbox{}ind
\begin{align}
 \log \la {\cal W}_1 {\cal W}_2 \ra_{\rm ILM} = nT\lambda_+ + {\rm
   (subleading\,\, terms)}.
\end{align}
In the last passage we have implicitly assumed that the projector
$\Pi_+$ is such that $  v_0 \cdot (\Pi_+ v_{N_c}) > 0$: 
we have explicitly verified that this is true in the cases that we
have investigated numerically. 
Note that the correlation function of parallel loops
must be positive, as a consequence of {\it ref\mbox{}lection positivity} of the 
Euclidean theory~\cite{OS}: this can be easily proved following
Ref.~\cite{Bachas}.  
Finally, recalling Eqs.~(\ref{eq:ddpot}) and (\ref{eq:ILM-qqbarpot}) 
we can conclude that
\begin{equation}
  {\cal V}_{dd}(\vec{d},\vec{R}_1,\vec{R}_2) = {\cal
    V}_{12}(\vec{d},\vec{R}_1,\vec{R}_2) - {\cal
    V}_{q\qbar}(|\vec{R}_1|)-{\cal V}_{q\qbar}(|\vec{R}_2|) ,
\end{equation}
where we have set ${\cal V}_{12} \equiv -n\lambda_+$.

In Figs.~\ref{fig:ddpot_pp}--\ref{fig:ddpot_apo} we show the comparison
of the instanton--induced {dipole--dipole potential} with some
preliminary numerical data obtained on the lattice: to our knowledge,
lattice measurements of this quantity are not present in the
literature. Also in this case we have performed a {\it quenched}
$SU(3)$ lattice calculation, with the same parameters ($16^4$
{hypercubic} lattice,
$\beta=6.0$) used in Ref.~\cite{lattice}, and so we have to use the
{\it quenched} instanton density $n_q$ and radius $\rho_q$, as explained in the
previous Section. The errors are very large for the
lattice data corresponding to the largest distances and lengths, and a
{\it plateau} has not been reached yet even at $T=0.8\,{\rm
  fm}$ at $|\vec{d}\,|=0\,{\rm fm}$ and $|\vec{d}\,|=0.1\,{\rm fm}$. However, we believe
that the order of 
magnitude will not change much for larger lengths: in the various
plots we show  the available values for ${\cal V}_{dd}$ as obtained
with loops of increasing lengths ($0.4\div 0.8\,{\rm fm}$), by means of
Eq.~(\ref{eq:ddpot}), and indeed on a logarithmic scale they seem to
be quite stable.  
The lattice value of ${\cal V}_{dd}$ extracted from the
largest--length loops is plotted at the correct value of the distance,
while the data points corresponding to shorter lengths are slightly
shifted, with the length increasing from left to right.
The instanton--induced potential and the lattice data have dif\mbox{}ferent
orders of magnitude for the shortest distances. At larger
distances there seems to be a better agreement, 
but the error bars of the lattice data are very large
there, and a better precision is needed to quantify the importance of 
instanton ef\mbox{}fects at large distances. However, as
already noticed in the previous Section, the rate of decrease with the
distance seems to be underestimated in the ILM, which would lead to an
increasing discrepancy between the prediction and the lattice data as
the distance between the dipoles increases.

\newsection{Conclusions}
\label{sec:inst_concl}

In this paper we have considered instanton ef\mbox{}fects,
in the framework of the ILM, on the Wilson--loop correlation function
relevant to {\it soft} high--energy scattering, and we have compared
the results to the lattice data discussed in Ref.~\cite{lattice}. 
{Using an approximate density function,} 
we have critically repeated the ILM calculation of the relevant
correlation function, 
already performed in Ref.~\cite{instanton1}, properly taking into account
ef\mbox{}fects which were neglected in that paper. 
The analytic dependence on the angular variable $\theta$ is the
same as the one found in Ref.~\cite{instanton1}, which we have used 
in Ref.~\cite{lattice}
to perform f\mbox{}its to the lattice data.
In this paper we have instead performed a direct comparison of the
ILM {\it quantitative} prediction, obtained by evaluating numerically our
analytic expression for the loop--loop correlation function in the
relevant cases. In doing so, we have used the pseudoparticle density
$n_q$ and radius $\rho_q$ appropriate for the {\it quenched}
approximation of QCD, in which the lattice calculation has been
performed. A direct numerical comparison allows us to test not only the given 
functional form, but also the quantitative relevance of instanton ef\mbox{}fects on
the Wilson--loop correlation function.

The comparison of the ILM prediction with the lattice 
data is not satisfactory, although it seems to have the correct
shape in the vicinity of $\theta=90^{\circ}$. 
Moreover, although the ILM prediction seems to be (more or less) of
the correct order of magnitude, the rate of decrease with the
transverse distance seems to be underestimated.  
Here one sees the importance of a {\it quantitative} comparison: terms
behaving as $1/\sin\theta$ (which are {\it qualitatively} good in
fitting the data around $\theta=90^{\circ}$) can have a different origin
than instantons, and indeed it seems that the ILM is not able to
explain them from a quantitative point of view.
Also, the ILM expression cannot account for the small but non--zero odderon
contribution to the correlation function which we have found in the lattice
results~\cite{lattice}. 
As a f\mbox{}inal remark, note that, after analytic continuation into Minkowski
space--time, the ILM result in the one--instanton approximation gives an exactly
zero total cross section, since the forward meson--meson scattering amplitude
turns out to be purely real. The situation can change after inclusion of
two--instanton ef\mbox{}fects, which are expected to yield a constant total cross
section at high energy~\cite{instanton1}; nevertheless, an analytic expression
for this contribution is currently unavailable.

In this paper we have also derived an analytic expression for the
instanton--induced dipole--dipole potential from the correlation function of
two parallel Wilson loops, which we have compared to some preliminary data
from the lattice.  
In this case, instanton ef\mbox{}fects seem to be negligible at short
distances, less than or equal to the size of the dipoles, where there is a
large discrepancy with the lattice results. The rate of decrease of the
potential with the spatial distance between the dipoles predicted by the ILM
seems also in this case to be smaller than the value found on the lattice;
a better accuracy in the numerical calculations is needed to make
this statement quantitative, but we think that the situation would not
change from a qualitative point of view.

\newpage

\appendix

\section{Dependence of the Wilson--loop correlation function on the
  longitudinal momentum fractions}
\label{app:longmom}

Exploiting the symmetries of the functional integral, one can 
show that the relevant Wilson--loop correlation function for given values of 
the longitudinal momentum fractions can always be reduced to the case
$f_1=f_2=1/2$. We discuss here the case of the physical, Minkowskian
correlation function ${\cal G}_M$; the proof in the Euclidean case is
perfectly analogous. 

As far as the dependence on the transverse vectors $\vec{z}_{\perp}$,
$\vec{R}_{1\perp}$, and $\vec{R}_{2\perp}$, and on the longitudinal momentum
fractions $f_1$ and $f_2$ is concerned, the invariance of the theory under
translations and spatial rotations implies that ${\cal G}_M$ can
depend only on the scalar 
products between the relative positions of quarks and antiquarks in transverse
space. The number of independent variables is then six, while
$\vec{z}_{\perp}$, $\vec{R}_{i\perp}$ and $f_i$ form a set of eight 
variables, so two of them are redundant. Indeed, the loop conf\mbox{}iguration
in the transverse plane is determined by the relative distances $\vec{R}_{1\perp}$
and $\vec{R}_{2\perp}$ between quarks and antiquarks inside each meson, and by
the relative distance between (say) quark 1 and quark 2 in the transverse
plane, 
\begin{equation}
  \vec{X}^{1q}_{\perp}(\tau)-\vec{X}^{2q}_{\perp}(\tau) = \vec{z}_{\perp} +
  (1-f_1)\vec{R}_{1\perp} - (1-f_2)\vec{R}_{2\perp}, \quad 
  \forall \tau.
\end{equation}
To determine which
conf\mbox{}igurations are equivalent, we have to solve the equation
\begin{align}
  & \vec{z}_\perp^{\,\prime} = \vec{z}_\perp +
  (f_1'-f_1)\vec{R}_{1\perp} - 
  (f_2'-f_2)\vec{R}_{2\perp}. 
\end{align}
Since a conf\mbox{}iguration specif\mbox{}ied by $(\vec{z}^{\,\prime}_\perp,f_1',f_2')$,
with $\vec{z}^{\,\prime}_\perp$ given above, is equivalent to the
conf\mbox{}iguration $(\vec{z}_\perp,f_1,f_2)$ we have 
\begin{align}
  {\cal
  G}_M(\chi;T;\vec{z}^{\,\prime}_\perp;\vec{R}_{1\perp},f_1',\vec{R}_{2\perp},f_2') =
  {\cal G}_M(\chi;T;\vec{z}_\perp;\vec{R}_{1\perp},f_1,\vec{R}_{2\perp},f_2); 
\end{align}
one can then choose a reference value for $f_i'$, for example $f_1'=f_2'=1/2$,
and move all the dependence on $f_i$ inside the dependence on the impact parameter.
Moreover, since ${\cal G}_M$ enters the dipole--dipole scattering amplitude with its 
2--dimensional Fourier transform with respect to the impact parameter, one can
write
\begin{align}
  \label{eq:nof}
  &\int d^2 z_{\perp}\, e^{i\vec{q}_{\perp}\cdot\vec{z}_{\perp}}  {\cal
    G}_M(\chi;T;\vec{z}_{\perp};\vec{R}_{1\perp},f_1,\vec{R}_{2\perp},f_2)\nonumber\\
  =&   \int d^2 z_{\perp}\, e^{i\vec{q}_{\perp}\cdot\vec{z}_{\perp}}  {\cal
  G}_M(\chi;T;\vec{z}_{\perp}-(f_1-\frac{1}{2})\vec{R}_{1\perp} +
    (f_2-\frac{1}{2})\vec{R}_{2\perp};\vec{R}_{1\perp},\frac{1}{2},\vec{R}_{2\perp},\frac{1}{2})
    \nonumber\\  
    =&
    e^{i\vec{q}_{\perp}\cdot[(f_1-\frac{1}{2})\vec{R}_{1\perp}-(f_2-\frac{1}{2})\vec{R}_{2\perp}]} 
    \int d^2 z_{\perp}\, e^{i\vec{q}_{\perp}\cdot\vec{z}_{\perp}}  {\cal
      G}_M(\chi;T;\vec{z}_{\perp};\vec{R}_{1\perp},\frac{1}{2},\vec{R}_{2\perp},\frac{1}{2}).
\end{align}
It is then clear that one can consider the case $f_1=f_2=1/2$ only.
In particular, in the forward case, i.e.,
$\vec{q}_{\perp}=0$, the dependence on $f_i$ drops, and the integration on the
longitudinal momentum fractions af\mbox{}fects only the wave functions.

\section{Technical details}

\label{app:C}

\setcounter{equation}{0}

\subsection{Integration over the position of pseudoparticles}

Consider the integral
\begin{equation}
 {\cal I} \equiv \int d^{4N}z\, {\cal D}(\{z\}) \prod_{k\in I_K}  f_k(z_{P(k)},\sigma_{P(k)}),
\end{equation}
where $f_k$ are $K$ (possibly distinct) matrices, whose entries
are functions of the position and of the type of the pseudoparticle lying
in cell $k$, and $I_K$ is the ordered set of the $K$ relevant cells. 
We have
\begin{equation}
\begin{aligned}
   {\cal I} = {\cal N}^{-1}\int d^{4N}z &\,\left[\sum_P\prod_{j=1}^{N}\chi_{D_j}(z_{P(j)})\right]
   \prod_{k\in I_K}  f_k(z_{P(k)},\sigma_{P(k)}) \\&= 
   {\cal N}^{-1}\sum_P V_{\rm cell}^{N-K} \int d^{4K}z \prod_{k\in I_K}
   \chi_{D_k}(z_{P(k)}) f_k(z_{P(k)},\sigma_{P(k)}) \\&=  
   {\cal N}^{-1}\sum_P V_{\rm cell}^{N-K}  \prod_{k\in I_K} \int_{D_k} d^{4}z\,
   f_k(z,\sigma_{P(k)}),
\end{aligned}
\end{equation}
which, denoting
\begin{equation}
  F_k(\sigma_{P(k)}) \equiv \int_{D_k} d^{4}z\, f_k(z,\sigma_{P(k)}),
\end{equation}
becomes
\begin{align}
\label{eq:inst-av-over-pos}
  {\cal I} &= {\cal N}^{-1}\sum_P V_{\rm cell}^{N-K}\prod_{k\in I_K}
  F_k(\sigma_{P(k)}) = {\cal N}^{-1}\sum_{S_K} V_{\rm cell}^{N-K}
  \eta(S_K)\prod_{k\in I_K}
  F_k(\sigma_{P(k)}),
\end{align}
where $\sum_{S_K}$ denotes the sum over all possible sequences of $K$
pseudoparticles, and $\eta(S_K)$ is the number of ways in which a
given sequence can be obtained from the ensemble,
\begin{equation}
   \eta(S_K) = (N-K)! \frac{N_{\inst}!}{(N_{\inst}-l_{\inst})!}
  \frac{N_{\ainst}!}{(N_{\ainst}-l_{\ainst})!},
\end{equation}
$l_{\inst}$ and $l_{\ainst}$ being the number of instantons and
anti--instantons in the sequence, respectively. In the limit of large
$N_{\inst}$, $N_{\ainst}$ and $V$, with the ratios $\nu_{\inst} =
N_{\inst}/N$, $\nu_{\ainst}=N_{\ainst}/N$, and $n=N/V$ kept f\mbox{}ixed,
\begin{equation}
   \eta(S_K) \to {\cal N} n^N \nu_{\inst}^{l_{\inst}}
   \nu_{\ainst}^{l_{\ainst}},
\end{equation}
and so Eq.~(\ref{eq:inst-av-over-pos}) simplif\mbox{}ies to (recall that $nV_{\rm
  cell} =1$)  
\begin{equation}
\begin{aligned}
\label{eq:inst-av-over-pos-2}
  {\cal I} &= n^K \sum_{S_K} \nu_{\inst}^{l_{\inst}}\nu_{\ainst}^{l_{\ainst}}
  \prod_{k\in I_K}  F_k(\sigma_{P(k)}) = n^K \sum_{S_K}
  \prod_{k\in I_K} \nu_{\sigma_{P(k)}} F_k(\sigma_{P(k)}) \\ &=
 n^K  \prod_{k\in I_K} \sum_{\sigma=\inst,\ainst} \nu_{\sigma} F_k(\sigma) =
  \prod_{k\in I_K} n\sum_{\sigma=\inst,\ainst} \nu_{\sigma}
\int_{D_k} d^{4}z\, f_k(z,\sigma). 
\end{aligned}
\end{equation}

\subsection{The function $B$ for $\nu_{\inst}=\nu_{\ainst}$}

We prove that the functions $B$ and $\widetilde{B}$, def\mbox{}ined in
Eq.~(\ref{eq:ddpotfunctions}), are equal for $\nu_{\inst}=\nu_{\ainst}$,
and f\mbox{}ind their explicit form. To do so, note the following about
$SU(2)$ matrices. One can always write ${\rm u}\in SU(2)$ as
\begin{equation}
  {\rm u} = u^0\I_2 + i\vec{u}\cdot\vec{\sigma},\quad u^0,u^i\in\mathbb{R},\quad
  (u^0)^2 + \vec{u}^{\,2} =1,
\end{equation}
with $u^0=\trd {\rm u}$ and $\vec{u} =-i\,\trd\vec{\sigma}{\rm u}$, with
$\trd \equiv (1/2)\tr$.
Using the commutation and anticommutation relations satisf\mbox{}ied by the
Pauli matrices,
\begin{align}
  [\sigma_i,\sigma_j] &= 2i\epsilon_{ijk}\sigma_k, \quad  \{\sigma_i,\sigma_j\} = 2\delta_{ij}\I_2,
\end{align}
we can write for the product of two unitary matrices
\begin{align}
  {\rm u} \equiv {\rm u}_1{\rm u}_2 = u^0\I_2 +i\vec{u}\cdot\vec{\sigma},
\end{align}
with
\begin{equation}
\begin{aligned}
u^0&=  [(u_1)^0(u_2)^0 - \vec{u}_1\cdot\vec{u}_2], &  
\vec{u}_{\phantom{a}} &= \vec{u}_s + \vec{u}_a,\\ 
  \vec{u}_s &= (u_1)^0\vec{u}_2 + (u_2)^0\vec{u}_1, & 
  \vec{u}_a &= -(\vec{u}_1\wedge\vec{u}_2).
\end{aligned}
\end{equation}
Clearly, if we interchange ${\rm u}_1$ and ${\rm u}_2$ we obtain
\begin{equation}
\begin{aligned}
  {\rm v} &\equiv {\rm u}_2{\rm u}_1 = {v}^0\I_2 +i\vec{v}\cdot\vec{\sigma},\\
  v^0 &= u^0,\quad  \vec{v} = \vec{u}_s - \vec{u}_a.
\end{aligned}
\end{equation}
In our case, denoting with 
${\rm w}_i^{(\pm)}(k)$ the non--trivial $2\times 2$ part of ${\rm
  W}^{(\pm)}_i(k)$, we have (suppressing the argument $k$)
\begin{equation}
  {\rm w}^{(\pm)}_i = w_{(\pm)i}^0\I_2 + \vec{w}_{(\pm)i}\cdot\vec{\sigma},
\end{equation}
so that
\begin{align}
{\rm w}_i\equiv   {\rm w}^{(-)\dag}_i{\rm w}^{(+)}_i &=
 w^0_i\I_2 +
  (\vec{w}_{i;s}+\vec{w}_{i;a})\cdot\vec{\sigma},
\end{align}
where, explicitly,
\begin{equation}
\begin{aligned}
  w^0_i &= [(w_{(-)i})^0(w_{(+)i})^0 +
    \vec{w}_{(-)i}\cdot\vec{w}_{(+)i}],\\ 
  \vec{w}_{i;s} &=
      [(w_{(-)i})^0\vec{w}_{(+)i}-(w_{(+)i})^0\vec{w}_{(-)i}],\\   
        \vec{w}_{i;a} &=
        \vec{w}_{(-)i}\wedge\vec{w}_{(+)i}\,.
\end{aligned}
\end{equation}
Since the vectors $\vec{w}_{(\pm)i}$ change sign if we replace an
instanton with an anti--instant\-on, 
we have that under this replacement $\vec{w}_{i;s}$ changes
sign, too, while $\vec{w}_{i;a}$ does not. Letting $\widetilde{\rm w}_1= {\rm
  w}^{(-)\dag}_2{\rm 
  w}^{(+)}_1$ and $\widetilde{\rm w}_2= {\rm w}^{(-)\dag}_1{\rm
  w}^{(+)}_2$, and $\bar{\rm w}_i = {\rm w}^{(+)}_i{\rm w}^{(-)\dag}_i$, we have 
\begin{equation}
\begin{aligned}
  1-\underline{w}_{12} &= \frac{2}{N_c}\{1-\trd[{\rm w}_1{\rm w}_2]\},\\
  1-\overline{w}_{12}  &= \frac{2}{N_c}\{1-\trd[\widetilde{\rm
        w}_1\widetilde{\rm w}_2]\}
 =  \frac{2}{N_c}\{1-\trd[\bar{\rm
        w}_1\bar{\rm w}_2]\};
\end{aligned}
\end{equation}
now,
\begin{equation}
\begin{aligned}
  \trd[{\rm w}_1{\rm w}_2] &= w^0_1w^0_2 -
  (\vec{w}_{1;s}+\vec{w}_{1;a})\cdot(\vec{w}_{2;s}+\vec{w}_{2;a}),\\
  \trd[\bar{\rm w}_1\bar{\rm w}_2]&= w^0_1w^0_2 -
  (\vec{w}_{1;s}-\vec{w}_{1;a})\cdot(\vec{w}_{2;s}-\vec{w}_{2;a}),
\end{aligned}
\end{equation}
and since we are averaging with equal weights over the pseudoparticle
species, we need to consider only terms which are symmetric under
$\inst\leftrightarrow\ainst$,
\begin{equation}
\begin{aligned}
  \trd[{\rm w}_1{\rm w}_2] &= w^0_1w^0_2 -
  (\vec{w}_{1;s}\cdot\vec{w}_{2;s}+ \vec{w}_{1;a}\cdot\vec{w}_{2;a}) +
  \,{\rm antisymmetric\,\,terms},\\
  \trd[\bar{\rm w}_1\bar{\rm w}_2]&= w^0_1w^0_2 -
  (\vec{w}_{1;s}\cdot\vec{w}_{2;s}+ \vec{w}_{1;a}\cdot\vec{w}_{2;a}) +
  \,{\rm antisymmetric\,\, terms}.
\end{aligned}
\end{equation}
From its def\mbox{}inition, Eq.~(\ref{eq:ddpotfunctions}), it is now
immediate to conclude that $B=\widetilde{B}$. Moreover, one can show
that  
\begin{equation}
\begin{aligned}
  &\vec{w}_{1;s}\cdot\vec{w}_{2;s}+ \vec{w}_{1;a}\cdot\vec{w}_{2;a} \\
  &=[(w_{(+)1})^0(w_{(+)2})^0 + \vec{w}_{(+)1}\cdot\vec{w}_{(+)2}]
  [(w_{(-)1})^0(w_{(-)2})^0 + \vec{w}_{(-)1}\cdot\vec{w}_{(-)2}]\\
&-
  [(w_{(-)1})^0(w_{(+)2})^0 + \vec{w}_{(-)1}\cdot\vec{w}_{(+)2}]
  [(w_{(-)2})^0(w_{(+)1})^0 + \vec{w}_{(-)2}\cdot\vec{w}_{(+)1}],
\end{aligned}
\end{equation}
which introducing
\begin{equation}
\begin{aligned}
  {\rm w}_+ &= {\rm w}_1^{(+)\dag}{\rm w}_2^{(+)} = {w}_+^0\I_2 +
  i\vec{ w}_+\cdot\vec{\sigma},\\ 
  {\rm w}_- &= {\rm w}_1^{(-)\dag}{\rm w}_2^{(-)} = {w}_-^0\I_2 +
  i\vec{ w}_-\cdot\vec{\sigma},
\end{aligned}
\end{equation}
leads to 
\begin{align}
  \trd[{\rm w}_1{\rm w}_2]|_{\rm symmetric} &= w^0_1w^0_2 +
  \widetilde{w}^0_1\widetilde{w}^0_2 - w^0_+w^0_- =
  \trd[\widetilde{\rm w}_1\widetilde{\rm w}_2]|_{\rm symmetric},
\end{align}
and thus to
\begin{equation}
\begin{aligned}
B = \int & d^3z\,\frac{1}{2}\big[\big(\underline{w}_{12}(z,\inst)-1\big)
  +\big(\underline{w}_{12}(z,\ainst)-1\big)\big] = \frac{2}{N_c}\int d^3z\,
  \big[ \big(\Delta_1\Delta_2 \\&
    +\widetilde{\Delta}_1\widetilde{\Delta}_2-\Delta_+\Delta_-\big)
    -  \big(\Delta_1 +\Delta_2 +\widetilde{\Delta}_1 +\widetilde{\Delta}_2
  -\Delta_+ - \Delta_-\big)\big] =\widetilde{B},
\end{aligned}
\end{equation}
with $\Delta_\pm$ given in the text, Eq.~(\ref{eq:deltapm}).

\newpage

\begin{table}[t]
  \centering
  \begin{tabular}{|c|ll|ll|ll|}
    \hline
        & \multicolumn{2}{c|}{predicted} &
    \multicolumn{2}{c|}{f\mbox{}itted--ILM} & \multicolumn{2}{c|}{f\mbox{}itted--ILMp}\\
    $|\vec{z}_{\perp}|\,[{\rm fm}]$ & $zzz$ & $zyy$  & $zzz$ & $zyy$ & $zzz$ & $zyy$ \\
    \hline
    0.0 
    &0.880--1.08 & 0.880--1.08
    &17.1 & 17.1                      
    &9.86 & 9.86 \\ 
    0.1 
    &0.827--1.02 & 0.798--0.984
    & 7.60 & 5.79 
    & 4.32 & 3.39\\
    0.2 
    &0.692--0.853 & 0.607--0.748
    &1.31 & 1.43
    &0.947 & 1.14 \\
    \hline
  \end{tabular}
\caption{Value of $K_{\rm ILM}\times 10^3$
  for the relevant conf\mbox{}igurations: 
  ILM prediction (f\mbox{}irst column), ``ILM'' f\mbox{}it (second column) and
  ``ILMp'' f\mbox{}it (third column).}
\label{tab:insttab}
\end{table}

\cleardoublepage

\cleardoublepage

\cleardoublepage

\cleardoublepage

\cleardoublepage

\cleardoublepage

\cleardoublepage

\cleardoublepage

\cleardoublepage

\cleardoublepage

\cleardoublepage

\cleardoublepage

\end{document}